\documentclass[11pt,preprint]{aastex}
\usepackage{psfig}
\def\gsim{\lower 2pt \hbox{$\, \buildrel {\scriptstyle >}\over
{\scriptstyle \sim}\,$}}
\def\lsim{\lower 2pt \hbox{$\, \buildrel {\scriptstyle <}\over
{\scriptstyle \sim}\,$}}

\def\chandra{{\sl Chandra~}}

\def\cor{\widehat=}

\def\xs{{NGC 3556}}

\shortauthors{}
\shorttitle{}

\begin{document}
\title{{\sl Chandra} Observation of the Edge-on Galaxy NGC 3556 (M~108):\\
Violent Galactic Disk-halo Interaction Revealed}
\author{Q. Daniel Wang}
\affil{Astronomy Department, University of Massachusetts, Amherst, MA 01003,USA}
\affil{Email: wqd@astro.umass.edu}
\author{Tara Chaves and Judith A. Irwin}
\affil{Department of Physics, Astronomy Research Group, Queen's University, Kingston, ON K7L 3N6, Canada}
\affil{tchaves@astro.queensu.ca and irwin@astro.queensu.ca}
\begin{abstract}

We present a 60 ks {\sl Chandra} ACIS-S observation of the isolated edge-on 
spiral NGC 3556, together with a multiwavelength analysis of various 
discrete X-ray sources and diffuse X-ray features. Among 33 discrete X-ray 
sources detected within the $I_B = 25$ mag arcsec$^{-2}$ isophote 
ellipse of the galaxy, we identify a candidate for the galactic 
nucleus, an ultraluminous X-ray source that might be an accreting 
intermediate-mass black hole, a possible X-ray binary with a radio 
counterpart, and two radio-bright giant HII regions. We detect large amounts 
of extraplanar diffuse X-ray emission, which extends about 10 kpc radially 
in the disk and
$\gtrsim 4$ kpc away from the galactic plane. The diffuse X-ray emission
exhibits significant substructures, possibly representing various blown-out 
superbubbles or chimneys of hot gas heated in massive star forming regions. 
This X-ray-emitting gas has temperatures in the range of $\sim 
2-7 \times 10^{6}$ K and has a total cooling rate of $\sim 2 \times 10^{40} 
{\rm~erg~s^{-1}}$. The energy can be easily supplied by supernova 
blast-waves in the galaxy. These results demonstrate NGC 3556 as being
a galaxy undergoing vigorous disk-halo interaction. The 
 halo in NGC~3556 is considerably less extended, however, than that
of NGC~4631, in spite of many similarities between the two galaxies.
This may be due to the fact that NGC~3556 is isolated whereas NGC~4631
is interacting.  Thus NGC~3556 presents a more pristine environment
for studying the disk-halo interaction.

\end{abstract}


\section{Introduction}

The galactic disk-halo interaction is believed to play an essential
role in the evolution of galaxies. The best way to investigate
the interaction is to observe nearby edge-on disk galaxies,
especially those ``normal'' ones, in which activity is not 
dominated by galactic nuclear regions. X-ray observations, in particular, 
are essential to the study of high-energy phenomena and processes that are 
likely responsible for much of the disk-halo interaction.
The \chandra {\it X-ray Observatory} with its arcsecond resolution
provides a uniquely effective tool to trace massive star
end-products, both stellar and interstellar (neutron stars, black holes,
and the hot interstellar medium) and allows for a clean separation
of point-like X-ray sources from diffuse emission, a step critical for 
reliably determining both the content and the physical state of 
diffuse hot gas.

The \chandra observation of the nearby edge-on spiral NGC~4631 (type Sd) 
provided the first unambiguous evidence for a hot gaseous galactic
corona around that galaxy (Wang et al. 2001). The overall X-ray emission from
the galaxy morphologically resembles the well-known radio halo of the
galaxy (e.g., as seen at 1.49 GHz; Hummel, Beck, \& Dahlem 1991; 
Wang et al. 1995), indicating a possible close link between outflows 
of hot gas and cosmic ray/magnetic field from the galactic disk. 
However, NGC~4631 may not be a ``typical'' normal disk galaxy because of its 
interaction with companions, as manifested by the galaxy's extended HI 
tidal tails. This interaction could have triggered the current 
high level of star formation as well as the outflows of the galaxy. It is
difficult to determine by studying NGC~4631 alone the relative importance of 
the external interaction and internal activity in producing 
the radio/X-ray-emitting halo. 

Here we present our \chandra observation of NGC 3556 (M108; Table 1) --- an 
isolated edge-on galaxy, which shows strong evidence for extraplanar 
radio emission (Fig. 1; e.g., Irwin et al. 1999). Similar to NGC~4631, 
this galaxy is also located in a region of low Galactic X-ray-absorbing 
gas column density, important for observing soft X-ray emission expected
from the galactic halo. Thus a comparison between these two galaxies allows
for probing the potential environmental effects on the structure of 
the disk/halo interaction.

\section{Data Calibration and Analysis}

Our \chandra observation of \xs~  (Ob. ID. 2025) was taken between Sept. 8-9, 
2001 for an exposure of 60 ks. The instrument ACIS-S was at the 
focal plane of the telescope. The galactic center was placed $\sim 1\farcm8$ 
off the telescope aim-point for an optimal coverage of the galaxy in
the S3 chip (Fig. 2). Although data from 
six chips (CCD ID \# 2, 3, 5, 6, 7, and 8)
were recorded, this work was based on the data from the latter three only, 
in particular the on-axis S3 (\# 7) chip (Fig. 2). 

We reprocessed the level 1 (raw) event data to generate a new level 2 event
file. This allowed us to capitalize on the latest 
Chandra Interactive Analysis of Observations software package
(CIAO; version 2.3), including an improved absolute astrometry
of the observation (nominally better than $\sim 0\farcs3$). We further removed time 
intervals with significant background flares --- peaks with count rates
$\gtrsim 3\sigma$ and/or a factor of 
$\gtrsim 1.2$ off the mean background level of the observation, 
using Maxim Markevitch's light-curve cleaning routine ``lc$\_$clean''
\footnote{available at http://hea-www.harvard.edu/~maxim/axaf/acisbg/}. 
This cleaning, together with a correction for the dead time of the 
observation, resulted in a net 58,111 s
exposure (livetime) for subsequent analysis.

Most of the remaining events are still not related to \xs, however.  A large 
portion of the events is due to charged cosmic-ray particles and to 
background/foreground X-rays of various origins.
The X-ray background/foreground varies from one part of sky to 
another. Particularly in the very soft band ($\lesssim 0.7$ keV), 
such variation is primarily due to 
the patchy contribution from the hot interstellar medium in our Galaxy.
The intensity of particle-induced events, which dominate 
in the energy range above 1.5 keV, varies with time, depending on solar 
activities and on the orbit location of the telescope. To estimate 
the average background spectrum and intensity, we used the so-called
blank-field data\footnote{available at 
http://cxc.harvard.edu/contrib/maxim/acisbg/data/README}. The data with
an effective livetime of 350 ks were re-projected to mimic our observation.
Unfortunately, our observation of \xs~ was taken in the `FAINT TIMED' mode,
whereas the blank-field data were a compilation of source-excised
observations taken in the ``VERY FAINT'' mode, which allowed
for a better software rejection of particle-induced events. For 
the back-illuminated chips (e.g., \# 7), the rejection could yield a 
significant background reduction at low or high 
energies, for examples, by  a factor of $\sim 5$ at $\sim 0.3$ keV or 
a factor of $\sim 1.3$ above 6 keV (A. Vikhlinin's 
online article\footnote{available at http://cxc.harvard.edu/cal/Acis/Cal$\_$prods/vfbkgrnd/index.html}). Therefore,
the use of the blank-field data underestimates 
particle-induced events in our observation.

For imaging analysis, we constructed exposure maps in four bands:
0.3-0.7, 0.7-1.5, 1.5-3, and 3-7 keV. These maps were used 
for flat-fielding, accounting for bad pixel removal as well as correcting for 
the telescope vignetting and the quantum efficiency
variation across the detector. We further corrected for the low energy
CCD quantum efficiency degradation with time\footnote{as discussed in the 
article available at 
http://cxc.harvard.edu/cal/Acis/Cal$\_$prods/qeDeg/index.html}. 
For our observation this degradation caused a fractional sensitivity 
reduction of 0.45 (or 0.14) in the 0.3-0.7 (or 0.7-1.5) keV band for 
the back-illuminated (BI) chips and a slightly smaller fraction
for the front-illuminated (FI) chips.
Accounting for the degradation is important for correctly 
converting a count
rate to an energy flux or comparing hardness ratios of an X-ray source 
with a spectral model (see later discussion). Our corrections assumed 
a power law spectrum with a photon index 
1.7 (a mean value for most known AGNs) as the weight within the individual 
bands. 

The effective energy-dependent spatial variation 
is weak within the S3 chip. A comparison 
between the 0.3-0.7 keV and  3-7 keV bands, for example, shows a 
relative variation of $\lesssim 15\%$. 
The energy dependency is stronger (up to a factor
of $\sim 3$) for multiple chips, because of different CCD 
types (BI vs. FI) and largely different off-axis locations. 
An intensity map is simply the background subtracted count map 
divided by the corresponding exposure map in each band. Combinations
of these maps can then be used to obtain intensity maps 
in broader energy bands (e.g., Figs. 1-3).

We searched for X-ray sources in three broad bands, 0.3-1.5 keV
(S), 1.5-7 keV (H), and
0.3-7 keV (B). A combination of source detection algorithms were applied: 
wavelet, sliding-box, and maximum likelihood centroid fitting.
First, we constructed the `Mexican cap'  wavelet images on
scales of 1, 2, 4 and 8 pixels (pixel size= 0\farcs492). 
On each scale we obtained a signal-to-noise (S/N) ratio map, which
is the ratio of 
the wavelet image and its uncertainty image. Local S/N maxima 
were detected as source candidates. 
Next, we applied a map detection (`sliding box' method) with a background 
map produced by removing sources detected with the wavelet 
method and smoothed adaptively to achieve a local count-to-noise ratio 
greater than 10. Finally, the sources detected with the map 
method were analyzed by a maximum likelihood algorithm, using both the 
background map and an approximate Gaussian point spread function 
corresponding to the respective off-axis angle. The estimation of the
count rate of a source was based on the data within the 90\% 
energy-encircled radius (EER) determined with the calibrated 
point spread function of the instrument, whereas the source removal 
adopted a larger source region (twice the 90\% EER).
Sources detected with a false detection probability $P\le 10^{-7}$ 
were finally accepted. 

For sources with count rates greater than $\sim 8 \times 10^{-3} 
{\rm~counts~s^{-1}}$, we conducted spectral fits with the XSPEC 
analysis software. We extracted a source spectrum from a radius 
that is twice the 90\% EER and a corresponding local background 
spectrum in a concentric annulus with the inner
and outer radii roughly 2.5 and 6 times the 90\% EER.
Sources within this annulus, if present, were removed.

Excluding the source regions (circles in Fig. 2, upper panel) from the data
allowed for the analysis of the  remaining ``diffuse'' 
emission from the galaxy. We estimated the X-ray intensity in the
source removed ``holes'', using  both event and exposure data 
in the surrounding bins (in swiss-cheese-like maps). This interpolation
resulted in a smoothed diffuse X-ray intensity map for visualization. 
But all quantitative measurements 
(surface brightness profiles, spectra, and luminosities) were based on
unsmoothed data. 

We analyzed the ACIS-S spectrum of the diffuse X-ray emission from
the galaxy. The on-galaxy
region (outlined by the dashed box in the upper panel of Fig. 2), 
covering essentially only the southeastern portion of the optical 
galaxy, was chosen to minimize the complication of X-ray absorption by the 
slightly tilted galactic disk on the northwestern part of the galaxy 
(see \S 3.2 for further discussion). For 
the background subtraction, we extracted an off-galaxy spectrum
from the two dash-dotted boxes of Fig. 2. To see whether or not this 
off-galaxy spectrum could actually be used to estimate the background 
contribution in the on-galaxy region, we compared the 
on-galaxy and off-galaxy spectra extracted from the simulated 
blank-field data. The spectral shapes and intensities of these two 
blank-field spectra were found to be statistically consistent
  with each other in the energy range of 0.3-7 keV, indicating no significant 
differences in the background property between the on- and off-galaxy regions.
Therefore, we used the off-galaxy spectrum from our data for the 
background subtraction of the on-galaxy spectrum. We further obtained weighted
effective area and response matrices, using the CIAO program ``ACISSPEC'' and
a diffuse X-ray weight map constructed in the 0.5-1.5 keV range to 
maximize the signal-to-noise ratio of the diffuse soft X-ray emission.

\section{Results}

\subsection{Discrete X-ray Sources}

Figs.~1 - 3 present complementary global views of NGC 3556, 
based on the \chandra data. Table 2 summarizes the results from our 
source detection. The note to this table explains various parameters listed.
The hardness ratios, in particular, provide simple source spectral
characteristics, which may be compared with spectral models (e.g., Fig. 4).
The detected sources are marked in Fig. 2, where a broad-band X-ray intensity 
image of the observation is presented together with an exposure map for 
ease of comparison. Fig. 3 presents a close-up of the galaxy. There is a 
clear concentration of X-ray sources, particularly ones with high count
rates, within the $R_{25}$ ellipse of \xs. A total of 33 detected sources 
are located inside the ellipse, which encloses a field of 10 arcmin$^2$. 
Our source detection is complete
down to a flux limit of $\sim 1 \times 10^{-15} {\rm~erg~s^{-1}~cm^{-2}}$ 
in the 2-10 keV band. The expected number of interlopers within the 
ellipse is only $\sim 7$ (e.g., Moretti et al. 2003). 
The following discussion concentrates
on bright sources within the ellipse and we will assume that they are located 
within the galaxy.

The source closest to the optical
centroid of \xs~ (Table 1) is J111131.3+554044 (Source 35 in Table 2). The kinematic center of the galaxy is $\alpha$ = $11^{\rm h} 11^{\rm m} 31\fs6 \pm 0\fs3$, $\delta$ = $55\degr 40^\prime 22^{\prime\prime} \pm 7\arcsec$ (King \& Irwin 1997) and coincides with the 
optical center of $\alpha$ = $11^{\rm h} 11^{\rm m} 30\fs97 \pm 1\farcs25$, $\delta$ = $55\degr 40^\prime 26\farcs8 \pm 1\farcs3$ from the 2MASS extended objects catalog (Jarrett et al. 2003).  The ACIS-S  spectrum of this 
source is too flat to be fitted with an one-temperature 
thermal plasma (XSPEC model ``MEKAL''; $\chi^2/{\rm d.o.f.} = 79/34$, where d.o.f. is degrees of freedom). Similarly, we
can also reject statistically ($\chi^2/{\rm d.o.f.} = 43.5/20$) a
multi-color disk black body (``DISKBB''), which is often 
used to characterize spectra of accreting disks around 
black holes in X-ray binaries. 
But a power law model, multiplied by a photon absorption model, gives a statistically satisfactory fit (Fig. 5; Table 3). 
The fitted high X-ray-absorbing gas column
density indicates that source is deeply embedded in the galaxy's disk. The 
0.5-10 keV luminosity is $\sim 2 \times 10^{39} 
{\rm~erg~~s^{-1}}$. Interestingly,
a plume-like diffuse X-ray emission feature seems to inflate from
the source and extend toward the south, more or less along
the minor axis of the galaxy (Fig. 3). Such
a diffuse X-ray feature is typically associated with an AGN or a nuclear
starburst. Therefore, source 35 with the second highest count rate in
Table 2 may represent the nucleus of \xs.

The source with the highest count rate is source 26 (Table 2), 
which is also projected in the 
disk of the galaxy (Fig. 3). A power law fit to the spectrum of this source 
can be rejected statistically at a confidence of $\sim 97\%$ 
($\chi^2/{\rm d.o.f.} = 72.6/55$). On the other hand, the disk blackbody model
fits well to the spectrum (Table 3, Fig. 5). The inner disk temperature of
$T_{in} \sim 1.3$ keV and the 0.5-10 keV luminosity of 
$\sim 4.6 \times 10^{39} {\rm~erg~s^{-1}}$ are well within the ranges
for so-called ultraluminous X-ray sources
(e.g., Makishima et al. 2000). 

Table 3 also includes the results from a satisfactory power law fit to 
the spectrum of source 25, with the 
third highest count rate in the field. The photon index
is characteristic of an AGN. The source is projected nearly 4$^\prime$ north of the 
center of NGC 3556, approximately on its minor axis, and is probably unrelated 
to the galaxy (Fig. 2, left panel).

A Kolmogorov-Smirnov test yields a positive variability detection of 
source 31, projected at $\sim 0\farcm5$ from 
the center of the galaxy (Fig. 3). 
We can statistically reject at a confidence greater than 3$\sigma$ the null 
hypothesis that the source flux was constant during the ACIS-S observation.
This variability was confirmed by a $\chi^2$ test (Fig. 6); the null hypothesis 
gives a $\chi^2$/d.o.f~=~56/7. The hardness ratios, HR and HR2, of the
source (as defined in the notes to Table 2) 
can be characterized by a power law spectrum of a photon index  
$\sim 2-3$ and a high absorption $\gtrsim 10^{21}
{\rm~cm^{-2}}$. The luminosity of the source is about a few $\times 10^{38} 
{\rm~erg~s^{-1}}$. Therefore, the source is likely to be an X-ray binary. 

For relatively faint X-ray sources, we have little spectral constraint.
A useful, though crude, conversion factor from a count rate to an
energy flux in the 0.3-7 keV band is  $\sim 6 \times  10^{-12}$ 
${\rm~(erg~cm^{-2}~s^{-1}})/({\rm counts~s^{-1}})$ for a relatively
hard X-ray source with a power law of a photon index $\lesssim 2$ or a 
plasma of a temperature of a few times $10^7$ K.
% The time-dependent 
%low energy degradation of the CCDs have approximately 
%been taken into account, an effect of $\sim 20\%$.
The conversion of the same 0.3-7 keV count rate to the 
2-10 keV flux is about a factor of 2 lower. These 
conversions are insensitive (within a factor $\sim 2$) to the absorbing gas 
column density, as long as it is smaller than a few times $\sim 10^{21} 
{\rm~cm^{-2}}$.
For soft X-ray sources (e.g., SNRs with a characteristic temperature 
$\lesssim 1$
keV), the conversion factor is a factor $\sim 2$ smaller for a column density 
 $\lesssim 10^{21} {\rm~cm^{-2}}$. In \S 4, we will further explore 
the nature of various X-ray 
sources by cross-correlating them with objects observed at other wavelengths. 

\subsection{Diffuse X-ray Emission}

Fig. 7 presents the intensity distribution 
of the diffuse X-ray emission from the NGC 3556 field. 
Relative to the major axis of the galaxy, the 
emission appears substantially more extended to the south than to the north. 
This is an indication of X-ray absorption by cool gas in
front of the X-ray-emitting region to the north, 
if the intrinsic distribution of the 
emission is more-or-less symmetric in respect to the major axis.
(Note that the galaxy's dust lane extends along the north side
of the disk.)
The projected emission region extends vertically (in the
minor axis direction) beyond the isophote
$I_B = 25$ mag arcsec$^{-2}$ ellipse, which defines an intrinsic 
disk radius $R_{25}$.
Along the major axis, however, the outer most intensity contour extends 
to only about 3/5 $R_{25}$ (``tangential'' points), which may be considered 
approximately as the radial extent of the diffuse X-ray emission region in the 
disk. At these ``tangential'' points, there should be little line-of-sight 
projection confusion. Therefore, the vertical height of X-ray emission 
at these points is most likely intrinsic. 

The diffuse emission shows substantial amounts of substructure. The 
lumpy morphology of the emission indicates that it does not
arise from a smoothed galactic corona in a hydrostatic equilibrium with the 
galaxy's gravity. There are apparent vertical X-ray ``spurs'', 
toward both the north and the south. Particularly at the east tangential 
points, such spurs extend $\gtrsim 1^\prime$ vertically from the major axis. 
This extent indicates a physical height $\gtrsim 4$ kpc. Some of the spurs 
might represent parts of larger-scale X-ray loops or rim-brightened 
blown-out bubbles, as indicated by 
faint diffuse X-ray structures in the halo of the galaxy (Fig. 7). However, 
deeper observations are required to draw any firm conclusion 
about these features. We further suspect that much of the diffuse emission 
from the galaxy may represent a composite of such discrete features. 

Fig. 8 compares the diffuse X-ray intensity distribution along 
the galaxy's minor axis and in different bands. 
In addition to the evident intensity enhancement associated with the galaxy,
there are apparent excesses 
(above the subtracted blank-field background) at distances
far away from the disk, in both the 0.3-0.7 keV and 1.5-7 keV bands. 
In particular, the intensity 
in the 0.3-0.7 keV band appears considerably stronger in the southeast than 
in the northwest of the disk. This 
difference may not be related to \xs\ and may instead represent an 
 intensity gradient in the Galactic foreground soft X-ray emission.  The excess seen in the harder 1.5-7 keV band is likely due to the expected, relatively high rate of particle-induced events in our observation (recall \S~2).

The diffuse emission intensity enhancement associated with \xs~
is predominantly in the 0.3-1.5 keV range and has a full width of about
2$^\prime$. The enhancement in the 0.3-0.7 band does not show a sharp peak 
at the galaxy's major axis as in the 0.7-1.5 keV band, but has a distribution
considerably broader than in the 0.7-1.5 keV band. The intensity enhancement 
also shows an apparent asymmetry in both bands, but it is 
particularly strong in the 0.3-0.7 keV band. This energy dependence 
suggests that the asymmetry is caused mainly by
soft X-ray absorption by the galactic disk, which presumably has its near
side tilted to the north. With the projected gas column density of 
$\gtrsim 2 \times 10^{21}{\rm~cm^{-2}}$ of the disk (King \& Irwin 1997), 
much of soft X-rays from the northern low halo or high disk are likely 
absorbed by the near-side part of the galaxy's disk. The emission from
the southern halo (southeast to the major axis) is not as much affected
by the disk.

The spectral properties of the diffuse X-ray
emission are characterized by simple-minded models available in XSPEC
(Fig. 9). The overall soft spectral characteristics 
indicate that the emission is primarily due to hot gas. 
However, an one-temperature, optically-thin, ionization equilibrium
plasma does not give a statistically 
acceptable fit ($\chi^2/{\rm d.o.f} =99.0/39$). A thermal model of two-temperature 
components, on the other hand, produces a statistically satisfactory fit,
not perfect though (Fig. 9; Table 4). There are apparent 
systematic deviations, especially an excess over the high energy part of 
the spectrum. This deviation is at least partly due to the presence of 
residual point-like sources in the disk of the galaxy. 
However, the counting statistics and spectral resolution of the present data
are too limited to be used for quantifying this plausible component.
The metal abundances are also not tightly constrained. Assuming 
independent abundances for the two components does not significantly
improve the fit.

\section{Multiwavelength Comparison}

We have correlated our X-ray source positions (Table 2) with six data sets 
from other wavelength bands of similar spatial resolution, including an
H$\alpha$ image taken with the {\sl Apache Point Observatory} 
(Collins et al. 2000), 
{\sl Two Micron All-Sky Survey} data, infrared data taken 
with the {\sl Infrared Space Observatory (ISO)}, {\sl Digitized Sky Survey (DSS)} data, archival images from the {\sl Hubble Space Telescope} WFPC2
(4678-7352 \AA, broad V-band), and a 20~cm radio continuum image taken with 
the {\sl Very Large Array} (Irwin et al. 2000). We concentrate our
discussion on the multiwavelength information that we have actually used to
delineate the nature of the observed X-ray emission from NGC 3556.

\subsection{Discrete Sources}

Fig. 10 shows an overlay of X-ray intensity contours on the 20 cm radio
image. We conducted an elliptical Gaussian fitting to measure 
the centroids and integrated emission intensities of radio 
objects (or peaks of extended emission) that show 
positional coincidences with X-ray sources. In Table 5, we list 
four such coincidences with the position
offsets between radio and X-ray centroids smaller than their combined 
3$\sigma$ position uncertainty radii. Of course, there is a chance that 
a positional coincidence may simply represent a random chance superposition.
Assuming both the radio and X-ray sources are randomly distributed
in the entire radio image, which is bigger than that shown in Fig. 10, we
estimate the probability to be $\sim 0.06$ or  
 $\sim 0.002$ for one or two chance superpositions. The realistic 
probability should be somewhat higher, however, because both radio and 
X-ray sources show a correlation 
with the galaxy. Nevertheless, most of the coincidences listed in Table 5
seem to represent real associations. 

We have explored the nature of the radio/X-ray position coincidences, 
based on their X-ray spectral 
characteristics. Source 18 has a hard X-ray spectrum,
as indicated by its hardness ratios (Table 2). Fig. 4 suggests that the
source is subject to heavy X-ray absorption ($\gtrsim 10^{21}{\rm~cm^{-2}}$). 
For a typical 
line-of-sight absorption in the disk of the galaxy, the spectral 
characteristics of the source are probably 
inconsistent with a thermal plasma with a temperature of a couple of keV, as 
expected for a young SNR, but appear to be consistent with
a power law  of a photon index $\sim 1$. 
The inferred luminosity of the source is $\sim 4 \times 10^{38} 
{\rm~erg~s^{-1}}$, for the distance of 
the galaxy (Table 1). The strong and extended radio emission (Table 5), 
however, is not expected for a typical X-ray binary. 
The only other suggested radio counterpart of a bright X-ray binary in 
nearby galaxies is an X-ray source with a luminosity 
of $\sim 1 \times 10^{40} {\rm~erg~s^{-1}}$ in the dwarf irregular 
galaxy NGC 5408 at the distance of 4.8 Mpc
(Kaaret et al. 2003). Even in this case, the radio flux is only 
$0.26 \pm 0.04$ mJy, a factor of $\sim 24$ lower than that inferred 
for source 18 here, accounting for the distance difference between the
two galaxies. Therefore,
the radio and X-ray association of source 19, if real, would be 
quite unusual.

Interestingly, the other three X-ray sources with radio counterparts in Table 5
are all best detected in the soft (S) band (Table 2). This means that the
X-ray fluxes of these sources are primarily in the energy range 
$\lesssim 1.5$ keV. In fact, these three are
the only such sources projected within the $R_{25}$ ellipse of the galaxy
(Fig. 3). There are three other such soft X-ray sources in Table 2 but they 
are all located far away from the galaxy (Fig. 1 left panel). 
Both source 24 and source 36 are relatively faint and
appear to be extended morphologically (Fig. 3). Therefore, these two sources
may represent giant HII regions such as the 30 Doradus nebula (Wang 1999).
Indeed, the substantially brighter radio counterpart of source 36
also appears correlated with objects seen in the H$\alpha$,
{\sl ISO} and {\sl HST} images.  

Source 53 has a luminosity of $\sim 3.5 \times 10^{38} {\rm~erg~s^{-1}}$ 
and is apparently point-like. The
offset from the radio-bright object (Table 5; Fig. 10) is greater than 
$2\sigma$. So the X-ray source and the radio object may not be related.
If physically associated, however, they could represent a very young SNR. 
Clearly, more information is needed to firmly establish the
nature of all these sources. For example, radio spectral indexes, as can be
afforded by a high-resolution radio observation at another wavelength, will
be particularly useful.

In order to compare the optical image with our X-ray data, two archival 
{\sl HST} images, both in the same V-band but at slightly different 
observing times, were first averaged together.  Next, we aligned 
three bright sources in the {\sl 2MASS} 
H-band image  with their counterparts in the {\sl HST} 
image.   Fig. 11 presents the combined {\sl HST} image which
covers the central region of the galaxy.  There are three X-ray sources, 
24, 36, and 42, which appear to coincide with optical sources within the 
astrometry offset error of $\sim 1^{\prime\prime}$.  Sources 24 and 
36 again appear correlated with bright, somewhat extended objects in 
the V-band.  
They have absolute magnitudes of $M_V = -13.3$ and $-14.0$, respectively.  
In a study of 145 HII regions in M33, Jiang
et al. (2002) found an average absolute magnitude in the V-band of -8.4.  Thus,
we are seeing emission in 3 bands, the 
X-ray, radio continuum and optical, from sources
 24 and 36, likely related to giant HII regions.
 Source 42 also appears 
correlated with a more point-like optical object and has a slightly fainter 
magnitude of $M_V = -11.7$
at the galaxy's distance. But the counting statistics of this X-ray source is 
too limited to provide useful constraints on its spectral properties.

\subsection{Diffuse Emission}

We compare the diffuse X-ray emission with observations in optical 
blue (Fig. 1, left panel), radio continuum (Fig. 1, right panel), and
H$\alpha$ line intensity (Fig. 12). Clearly, there is 
a good overall morphological similarity between the radio and X-ray emissions.
The elongation along the galactic disk is in sharp contrast to the vertical 
orientation (along the minor axis) typically seen in the diffuse X-ray 
emission from a nuclear starburst galaxy (e.g., M82 and NGC 253). 
While the vertical morphology is a sign of galactic nuclear 
superwind, the large-scale disk-oriented extraplanar diffuse 
X-ray emission from 
NGC~3556 suggests an active disk/halo interaction. The limited vertical
extent of the emission probably indicates that the hot gas is
confined by the gravity of the galaxy. 

From the spectral parameters in Table 4, we may estimate the
physical parameters of the diffuse X-ray-emitting gas. 
We assume that the gas is located within 
a cylinder with its axis aligned with the rotation axis of the galaxy.
This cylinder has an angular radius 2\farcm6 (3/5$R_{25}$) and a height 2$^\prime$,
which gives a total volume of $\sim 9.0 \times 10^{67}{\rm~cm^{3}}$.
The best-fit integrated emission measures (Table 4), 
multiplied by a factor of 2
to account for the contribution from the northern halo, suggest
a mean electron density of $\sim 2.3 \times 10^{-3}{\rm~cm^{-3}} 
\xi_l^{-1/2}$ and $1.6 \times 10^{-3}{\rm~cm^{-3}} 
\xi_h^{-1/2}$ for the low and high temperature components with the
volume filling factors, $\xi_l$ and $\xi_h$, respectively. The total
volume filling factor is $\xi=\xi_h + \xi_l$. Assuming that there is 
a thermal pressure balance between the two components, we find
$\xi_h \sim 0.7\xi$. We further infer the mean pressure and total thermal energy 
of the hot gas as $\sim 2.6 \times 10^4 {\rm~K~cm^{-3}}\xi^{-1/2}$ 
 and $\sim 5
\times 10^{56} {\rm~erg~}\xi^{1/2}$. The total cooling rate 
of the gas is $\sim 2 \times 10^{40} {\rm~ergs~s^{-1}}$.
The cooling time of the gas is $\sim 8 \times 10^8\xi^{1/2}$ years. 

It is interesting to compare the thermal pressure of the hot gas
(equivalent to $\sim 3.5 \times 10^{-12}~{\rm dyne~cm^{-2}}$
$\xi^{-1/2}$ 
from above)  with
the magnetic field pressure in the galaxy. Irwin et al. (1999)
show resolved maps of the magnetic field strength in NGC~3556
using the minimum energy assumption.
 The mean magnetic field, adjusting
to the currently adopted distance, is 6.7 $\mu$G, with a range from
5.1 to 10.2 $\mu$G.  The magnetic field pressure, assuming
isotropy
and uniform filling, 
is 1.8 $\times$ 10$^{-12}$ dyne cm$^{-2}$ (ranging from
1.0 to 4.1 $\times$ 10$^{-12}$ dyne cm$^{-2}$).  
%Thus, the magnetic field pressure appears to be lower than the thermal 
% pressure by a factor of $\sim$ 4, and by more if the filling factor for
% the thermal gas is less than unity or less than the filling
%factor for the magnetic field.  This indeed seems likely since   
This, together with the accompanied contribution of cosmic-rays, may be 
comparable to the thermal pressure at the galactic disk.
The projected region over which
the magnetic field has been determined is actually smaller than the
region over which the thermal pressure has been determined (see
Fig. 12 of Irwin et al.), i.e. the magnetic pressure calculation
applies to a region more confined to the disk. Therefore, a more realistic 
comparison
must account for the pressure variation in both the radio- and X-ray-emitting
materials.
%Since we do not
%expect magnetic pressure to increase away from the plane, it is
%likely that thermal pressure dominates magnetic pressure in the halo region.

What is the origin of the X-ray-emitting gas? Fig. 12 shows some correlation
between the diffuse X-ray and H$\alpha$ emissions,
indicating that massive stars are responsible for the heating of the gas
in the galactic disk. Massive stars tend to form in clusters or associations.
The concentrated mechanical energy input in the form of fast stellar winds and
supernovae (SNe) can naturally generate superbubbles of hot gas. Such superbubbles,
if energetic, can break out from the disk into the halo, which 
may explain the various X-ray ``spurs'' observed. There do appear to
be some high latitude HI extensions (from data in King \& Irwin 1997)
associated with X-ray spurs, but the spatial scales are not similar
enough to draw firm conclusions.
It would be interesting
to have a detailed study of the correlation between optical and X-ray 
emission structures, although the optical appearance of these structures
may be affected severely by the extinction/absorption  in the disk. 
Dust lanes are clearly seen in
the northern part of the optical disk and appear to coincide with
regions of relatively low diffuse X-ray intensity or high intensity
gradient (Fig. 12). Similar effects can also been seen on larger 
scales in Fig. 1 (left panel).

 NGC 3556 is an isolated galaxy (e.g., Davis \& Seaquist 1983). It is 
thus unlikely that the HI supershells observed in this galaxy are produced from 
impacting clouds (Irwin et al. 1999). As seen in other galaxies,
some such shells are clearly associated with massive 
star forming regions, as traced by \ion{H}{2} regions. Thus, the energy 
may be provided by SNe and stellar winds from massive stars. 
Our estimated 
X-ray-emitting gas cooling rate
corresponds to a SN rate of $\sim 6$ per 1000 years, assuming 
a thermal energy input of $\sim 10^{50} {\rm~erg~s^{-1}}$ per SN.  The SN rate for \xs~ is $\sim 32$ per 1000 years (Irwin, English \& Sorathia, 1999).
Therefore, the thermal energy of the diffuse hot gas observed could be 
easily supplied by SNe in the disk of the galaxy.  We caution, however,
that there are other very large and energetic HI extensions in NGC~3556
which do not seem to be explained via underlying star formation (King \&
Irwin 1997), so although there may be a connection between 
massive star formation
and X-ray spurs in the inner galaxy, star formation alone may not
explain all of the energetic features in this galaxy.

Finally, let us compare the results on \xs~ and NGC 4631.
Both galaxies show a similar overall morphology of extraplanar emission 
in radio continuum and diffuse X-ray, indicating that there is a
close link between the outflows of hot gas and cosmic-rays/magnetic fields
in galaxies (Wang et al. 1995 and reference therein). 
 Furthermore, H$\alpha$ images of NGC~4631 show a 
prominent ``double worm'' structure plus a forest of faint vertical filaments
(or loops) emanating from the galactic disk of NGC 4631 toward its
northern halo, where the extraplanar emission is most extended (Rand et al.
1992; Wang et al. 2001).  This again argues for the extraplanar emission
to be linked to underlying processes.  Furthermore, NGC~4631 and NGC~3556 have 
nearly identical
ratios of the total far-infrared flux to the blue light flux as well as
dust temperatures (Rice et al. 1988), indicating comparable star formation
rates in  the two galaxies. If we wish to consider
how concentrated the star formation is in the two galaxies, the
best indicator for edge-on dusty systems is the radio continuum
emission.  
NGC~3556 shows an increase in its radio continuum emission within
its central 1\farcm4 (5.9 kpc; Irwin et al. 1999) indicative
of a centrally enhanced region of star formation.  Similarly,
NGC~4631 (7.6 Mpc) also shows increasing radio continuum emission
within the central $\sim$ 3 to 3\farcm5 (6.6 to 7.7 kpc; e.g., Dumke et al. 1995).  Both galaxies also have
rotation curves that peak at
$\sim$ 150~km~s$^{-1}$, relative to their systemic velocities 
(Rand 1994; King \& Irwin 1997) indicating that they have similar
mass and therefore similar abilities to retain gas that has been
ejected into the halo.
Yet the emission in both radio and X-ray is considerably more
extended in NGC 4631 along the minor axis than in \xs. What might be
the cause of this difference?  At the present time, the only obvious
difference between \xs~ and NGC~4631 is that the latter is involved in
a galaxy-galaxy interaction whereas the former is isolated.  
A larger
sample of galaxies is required to see whether galaxy halos are more
prominent in interacting systems, but this result is interesting. 
 \xs~ probably
presents a cleaner case for studying the disk-halo interaction driven by 
galactic internal processes than interacting galaxies like NGC~4631.

\section{Summary}

We have presented a detailed analysis of the \chandra ACIS-S observation
of the edge-on galaxy 
NGC 3556. Together with existing observations at other wavelengths, we
have explored the nature and physical state of various X-ray sources and 
features as well as large-scale diffuse soft X-ray emission from the galaxy.
Our main results and conclusions are as follows:

\begin{itemize}

\item We have detected 83 X-ray sources in the field of NGC 3556 (Table 2; Fig. 2). 
Among them, 33 are within the R$_{25}$ ellipse of the galaxy. 
One source shows strong timing variability during the observation.

\item We have identified source 35 (J111131.3+554044) as 
a candidate for the nucleus of the galaxy. This source, with
the second highest count flux detected here, has a power law
X-ray spectrum typical for an AGN (Table 3). 
In addition, the source is possibly
associated with an X-ray-emitting plume pointing toward the southern halo.

\item The brightest X-ray source detected (Source 26), 
with a 0.5-10 keV luminosity of 
$\sim 5 \times 10^{39} {\rm~erg~~s^{-1}}$, has a spectrum well 
characterized by a multi-color disk blackbody (Table 3). This 
ultraluminous X-ray source
may thus represent an accreting intermediate-mass black hole. 

\item We have found possible radio counterparts for four X-ray sources
(Table 5). One of them  (Source 18)
has a hard X-ray spectrum and may thus be an X-ray binary. The radio
counterpart, however, is unusually bright and extended.
The other three sources are the only ones detected primarily in the 
0.3-1.5 keV band and within the R$_{25}$ ellipse. Two of them 
(Sources 24 and 36) have extended and relatively faint counterparts in the radio and extended and relatively bright counterparts in the optical, most likely representing hot gas in giant HII 
regions.

\item We have detected large amounts of diffuse soft X-ray emission 
from the galaxy (e.g., Fig. 7). This emission is confined within a radius of 
$\sim 10$ kpc (3/5$R_{25}$) in the galactic disk, but extends beyond $\sim 4$ 
kpc into the halo of the galaxy. The overall morphology of the emission 
suggests an active disk-halo interaction throughout much of the galaxy.

\item The diffuse X-ray emission morphology shows a significant 
asymmetry relative to the major axis of
the galaxy. This is most likely due to the X-ray absorption by the nearly 
edge-on galactic disk that is tilted with the northern side closer to us 
than the southern one.

\item The diffuse X-ray emission shows substructures such as vertical ``spurs'', 
which may represent breaking out superbubbles from the galactic disk. 
There is a good spatial correlation between diffuse X-ray emission and
H$\alpha$ features within the disk, indicating that 
X-ray-emitting gas originates primarily in recent massive star forming regions.

\item The spectrum of the diffuse emission can be characterized approximately
by a two-temperature thermal plasma with a total luminosity of 
$\sim 2 \times 10^{40} {\rm~erg~s^{-1}}$. This energy loss is likely balanced
by SN heating in the disk. 

\item In spite of many similarities between NGC~3556 and NGC~4631, the halo
of NGC~4631 is much more extended than in NGC~3556.  This may be due to
the fact that NGC~4631 is interacting whereas NGC~3556 is isolated and 
therefore offers
a more pristine environment for studying disk-halo interactions.

\end{itemize}

In short, this study demonstrates that \xs~ is a fine example for a wealth of
high-energy phenomena and processes in an active starforming disk
galaxy. In particular, the extensive extraplanar diffuse X-ray emission 
and its substructure revealed here 
are clearly a manifestation of the intense disk-halo interaction in NGC 3556.

\acknowledgements
We thank Melissa Ruiters for helping us with IRAF, Dave Smith for helping in
the X-ray data calibration, Sally Oey and Eric Schlegel for helpful comments. 
This work was funded partly by NASA under the grants 
GO1-2084A and NAG5--8999.  
\vfil
\eject

\vfil
\eject
%%============================
%% Tables 
\begin{deluxetable}{lrr}
\tabletypesize{\footnotesize}
\tablecaption{Salient Parameters of NGC~3556 \label{parameters}}
\tablewidth{0pt}
\tablehead{
\colhead{Parameter} &
\colhead{Value} & 
\colhead{Ref.}}
\startdata
Type     \dotfill & SB(s)cd & 1 \\
$R_{25}$\dotfill & 4\farcm35 & 1 \\
Inclination angle  \dotfill & $\sim 80^\circ$& 1 \\
Position  angle  \dotfill & $\sim 80^\circ$& 1 \\
Center position \dotfill & 
        R.A.~$11^{\rm h} 11^{\rm m} 30\fs97$ & 2 \\
~~ (J2000)\dotfill & Dec.~$55\degr 40^\prime 26\farcs8$ & 2 \\
Blue Magnitude  \dotfill &  10.7 mag & 1\\
Distance \dotfill & $14.1$~Mpc & 3 \\
\dotfill & ($1^\prime~\cor~4.1$~kpc) & \\
Galactic foreground $N_{\rm HI}$ \dotfill & 
$1.3 \times 10^{20}$~cm$^{-2}$ & 4 \\
\enddata
\tablerefs{
(1) NED;
(2) Jarrett et al. (2003)
(3) Ho et al. (1997) and references therein;
(4) Dickey \& Lockman (1990)
}
\end{deluxetable}
\vfill
\eject
%====================================================
%Table 2
%====================================================
%%
%====================================================
%Table 1
\begin{deluxetable}{lrrrrrrrrrr}
  \tabletypesize{\footnotesize}
  \tablecaption{{\sl Chandra} Source List \label{acis_source_list}}
  \tablewidth{0pt}
  \tablehead{
  \colhead{Source} &
  \colhead{CXOU Name} &
  \colhead{$\delta_x$ ($''$)} &
  \colhead{log(P)} &
  \colhead{CR $({\rm~cts~ks}^{-1})$} &
  \colhead{HR} &
  \colhead{HR1} &
  \colhead{HR2} &
  \colhead{Flag} \\
  \noalign{\smallskip}
  \colhead{(1)} &
  \colhead{(2)} &
  \colhead{(3)} &
  \colhead{(4)} &
  \colhead{(5)} &
  \colhead{(6)} &
  \colhead{(7)} &
  \colhead{(8)} &
 \colhead{(9)}
  }
  \startdata
   1 &  J111054.09+554113.2 &  1.6 &$  -8.5$&$     0.49  \pm   0.22$&                                       --& --& -- &  B \\
   2 &  J111101.25+554049.5 &  0.6 &$ -20.0$&$     2.01  \pm   0.25$&             $-0.56\pm0.14$ & $ 0.12\pm0.15$ & -- &  B \\
   3 &  J111106.00+554416.7 &  1.1 &$ -14.7$&$     0.65  \pm   0.14$&                                       --& --& -- &  B \\
   4 &  J111108.15+554151.4 &  0.7 &$ -15.7$&$     0.83  \pm   0.16$&                                       --& --& -- &  B \\
   5 &  J111108.22+554013.5 &  1.4 &$  -8.6$&$     0.32  \pm   0.10$&                                       --& --& -- &  B \\
   6 &  J111108.50+554245.6 &  0.7 &$ -20.0$&$     1.17  \pm   0.19$&                                       --& --& -- &  B \\
   7 &  J111109.37+554248.7 &  0.4 &$ -20.0$&$     1.62  \pm   0.20$&  $ 0.07\pm0.17$ & $ 0.69\pm0.17$ & $-0.19\pm0.16$ & B \\
   8 &  J111109.77+553959.7 &  0.6 &$ -20.0$&$     0.69  \pm   0.15$&                                       --& --& -- &  B \\
   9 &  J111110.67+553950.5 &  1.0 &$  -8.9$&$     0.31  \pm   0.10$&                                       --& --& -- &  B \\
  10 &  J111112.85+553759.9 &  0.7 &$ -20.0$&$     0.99  \pm   0.18$&                                       --& --& -- &  B \\
  11 &  J111113.54+554016.0 &  0.5 &$ -20.0$&$     1.09  \pm   0.16$&               --& $ 0.84\pm0.17$ & $-0.62\pm0.17$ & B \\
  12 &  J111114.82+554416.2 &  1.3 &$  -8.1$&$     0.35  \pm   0.11$&                                       --& --& -- &  B \\
  13 &  J111115.41+554002.8 &  0.7 &$  -8.3$&$     0.37  \pm   0.11$&                                       --& --& -- &  B \\
  14 &  J111116.36+554142.4 &  0.3 &$ -20.0$&$     2.23  \pm   0.25$&             $-0.59\pm0.12$ & $ 0.03\pm0.13$ & -- &  B \\
  15 &  J111116.55+554135.8 &  0.4 &$ -15.6$&$     1.33  \pm   0.19$&             $-0.44\pm0.17$ & $ 0.16\pm0.18$ & -- &  B \\
  16 &  J111116.66+554333.8 &  1.0 &$  -9.2$&$     0.33  \pm   0.10$&                                       --& --& -- &  B \\
  17 &  J111117.70+554009.9 &  0.2 &$ -20.0$&$     5.86  \pm   0.39$&  $-0.42\pm0.08$ & $ 0.29\pm0.08$ & $-0.72\pm0.08$ & B \\
  18 &  J111117.80+554016.7 &  0.2 &$ -16.3$&$     2.75  \pm   0.24$&  $ 0.33\pm0.11$ & $ 1.00\pm0.08$ & $-0.01\pm0.10$ & B \\
  19 &  J111118.69+554238.1 &  1.1 &$ -14.4$&$     0.43  \pm   0.12$&                                       --& --& -- &  S \\
  20 &  J111119.13+553835.4 &  0.5 &$ -16.1$&$     0.76  \pm   0.14$&                                       --& --& -- &  B \\
  21 &  J111120.62+554103.1 &  0.6 &$ -12.4$&$     0.38  \pm   0.10$&                                       --& --& -- &  B \\
  22 &  J111121.94+554016.4 &  0.5 &$ -20.0$&$     0.49  \pm   0.11$&                            --& --& $ 0.58\pm0.17$ & B \\
  23 &  J111122.76+554030.4 &  0.4 &$ -16.3$&$     0.74  \pm   0.13$&                                       --& --& -- &  B \\
  24 &  J111124.70+554041.8 &  0.4 &$ -20.0$&$     0.45  \pm   0.12$&                                       --& --& -- &  S \\
  25 &  J111125.72+554401.4 &  0.2 &$ -15.7$&$     8.75  \pm   0.48$&  $-0.49\pm0.06$ & $ 0.06\pm0.07$ & $-0.38\pm0.09$ & B \\
  26 &  J111126.02+554016.7 &  0.1 &$ -20.0$&$    26.77  \pm   0.77$&  $-0.03\pm0.04$ & $ 0.69\pm0.04$ & $-0.33\pm0.04$ & B \\
  27 &  J111126.13+553945.1 &  0.6 &$  -7.4$&$     0.20  \pm   0.08$&                                       --& --& -- &  B \\
  28 &  J111126.80+554028.6 &  0.2 &$ -15.8$&$     2.14  \pm   0.23$&  $-0.38\pm0.13$ & $ 0.63\pm0.12$ & $-0.49\pm0.15$ & B \\
  29 &  J111126.87+553914.7 &  0.2 &$ -16.0$&$     5.76  \pm   0.38$&  $-0.44\pm0.08$ & $ 0.35\pm0.08$ & $-0.26\pm0.11$ & B \\
  30 &  J111127.23+554004.9 &  0.6 &$  -8.6$&$     0.28  \pm   0.09$&                                       --& --& -- &  B \\
  31 &  J111128.30+554021.2 &  0.3 &$ -15.9$&$     1.21  \pm   0.18$&               $ 0.24\pm0.20$ & --& $-0.10\pm0.18$ & B \\
  32 &  J111129.23+553621.2 &  1.1 &$  -9.4$&$     0.32  \pm   0.11$&                                       --& --& -- &  B \\
  33 &  J111130.05+554011.6 &  0.4 &$ -10.7$&$     0.39  \pm   0.11$&                                       --& --& -- &  B \\
  34 &  J111130.23+553629.9 &  0.6 &$ -20.0$&$     0.71  \pm   0.14$&                                       --& --& -- &  B \\
  35 &  J111130.32+554031.1 &  0.1 &$ -20.0$&$     8.56  \pm   0.49$&  $-0.08\pm0.08$ & $ 0.72\pm0.07$ & $-0.23\pm0.08$ & B \\
  36 &  J111131.34+554044.0 &  0.3 &$ -15.1$&$     0.57  \pm   0.13$&                                       --& --& -- &  S \\
  37 &  J111131.61+554041.0 &  0.4 &$  -9.0$&$     0.34  \pm   0.10$&                                       --& --& -- &  B \\
  38 &  J111131.80+553931.6 &  0.6 &$ -10.4$&$     0.25  \pm   0.08$&                                       --& --& -- &  B \\
  39 &  J111131.81+554041.3 &  0.4 &$  -8.8$&$     0.33  \pm   0.10$&                                       --& --& -- &  B \\
  40 &  J111132.39+554027.0 &  0.3 &$ -12.7$&$     0.47  \pm   0.12$&                                       --& --& -- &  B \\
  41 &  J111132.63+554405.0 &  0.8 &$ -11.0$&$     0.31  \pm   0.09$&                                       --& --& -- &  S \\
  42 &  J111132.64+554032.8 &  0.3 &$ -16.3$&$     0.90  \pm   0.14$&                          --& $ 0.92\pm0.13$ & -- &  B \\
  43 &  J111132.91+554032.0 &  0.3 &$ -20.0$&$     0.79  \pm   0.13$&               $ 0.86\pm0.19$ & --& $ 0.21\pm0.16$ & B \\
  44 &  J111134.35+554149.1 &  0.5 &$ -20.0$&$     0.35  \pm   0.09$&                                       --& --& -- &  B \\
  45 &  J111134.53+554249.8 &  0.2 &$ -16.6$&$     2.28  \pm   0.25$&  $-0.44\pm0.13$ & $-0.00\pm0.13$ & $-0.42\pm0.16$ & B \\
  46 &  J111135.24+554107.2 &  0.2 &$ -16.0$&$     2.18  \pm   0.23$&  $-0.19\pm0.14$ & $ 0.51\pm0.14$ & $-0.53\pm0.13$ & B \\
  47 &  J111135.54+554222.8 &  0.5 &$ -20.0$&$     0.47  \pm   0.11$&                                       --& --& -- &  B \\
  48 &  J111136.94+553929.5 &  0.9 &$  -9.8$&$     0.17  \pm   0.07$&                                       --& --& -- &  H \\
  49 &  J111137.80+553728.3 &  0.8 &$  -8.0$&$     0.19  \pm   0.08$&                                       --& --& -- &  H \\
  50 &  J111138.13+554019.1 &  0.1 &$ -17.1$&$     3.95  \pm   0.31$&  $-0.45\pm0.09$ & $ 0.52\pm0.09$ & $-0.65\pm0.10$ & B \\
  51 &  J111138.19+554236.7 &  0.1 &$ -20.0$&$     4.90  \pm   0.32$&  $ 0.59\pm0.07$ & $ 1.00\pm0.07$ & $ 0.09\pm0.07$ & B \\
  52 &  J111138.37+553951.6 &  0.1 &$ -15.9$&$     2.74  \pm   0.26$&  $-0.34\pm0.12$ & $ 0.45\pm0.12$ & $-0.10\pm0.15$ & B \\
  53 &  J111138.99+554102.5 &  0.1 &$ -20.0$&$     2.64  \pm   0.25$&               --& $ 0.74\pm0.08$ & $-1.00\pm0.19$ & S \\
  54 &  J111139.05+553955.5 &  0.4 &$ -13.3$&$     0.27  \pm   0.09$&                                       --& --& -- &  B \\
  55 &  J111139.40+553954.1 &  0.2 &$ -16.1$&$     0.79  \pm   0.13$&                          --& $ 0.86\pm0.16$ & -- &  B \\
  56 &  J111140.32+554025.5 &  0.2 &$ -16.6$&$     0.67  \pm   0.13$&                                       --& --& -- &  B \\
  57 &  J111140.48+554012.5 &  0.1 &$ -20.0$&$     2.72  \pm   0.25$&  $-0.32\pm0.12$ & $ 0.65\pm0.11$ & $-0.46\pm0.13$ & B \\
  58 &  J111141.39+554057.7 &  0.1 &$ -20.0$&$     4.87  \pm   0.32$&  $ 0.52\pm0.07$ & $ 1.00\pm0.06$ & $-0.28\pm0.07$ & B \\
  59 &  J111143.34+554221.9 &  0.4 &$ -13.8$&$     0.28  \pm   0.09$&                                       --& --& -- &  B \\
  60 &  J111143.76+554246.9 &  0.3 &$ -20.0$&$     0.85  \pm   0.15$&                                       --& --& -- &  S \\
  61 &  J111144.77+554335.8 &  0.4 &$ -16.3$&$     0.71  \pm   0.13$&                                       --& --& -- &  B \\
  62 &  J111145.05+554012.7 &  0.5 &$ -12.2$&$     0.23  \pm   0.08$&                                       --& --& -- &  B \\
  63 &  J111145.74+553857.0 &  0.3 &$ -20.0$&$     0.38  \pm   0.09$&                                       --& --& -- &  B \\
  64 &  J111146.69+553943.6 &  0.6 &$  -9.9$&$     0.18  \pm   0.07$&                                       --& --& -- &  B \\
  65 &  J111147.13+553843.9 &  0.3 &$ -20.0$&$     0.35  \pm   0.10$&                                       --& --& -- &  B \\
  66 &  J111148.98+554024.4 &  0.1 &$ -16.2$&$     1.43  \pm   0.20$&             $-0.54\pm0.14$ & $ 0.10\pm0.16$ & -- &  B \\
  67 &  J111150.41+554223.0 &  0.3 &$ -16.0$&$     0.56  \pm   0.12$&                                       --& --& -- &  B \\
  68 &  J111152.39+553920.0 &  0.4 &$ -10.9$&$     0.20  \pm   0.08$&                                       --& --& -- &  B \\
  69 &  J111153.96+553728.2 &  0.3 &$ -20.0$&$     1.88  \pm   0.22$&  $-0.20\pm0.15$ & $ 0.51\pm0.15$ & $-0.35\pm0.16$ & B \\
  70 &  J111155.31+553846.7 &  0.2 &$ -16.1$&$     1.32  \pm   0.18$&  $-0.25\pm0.18$ & $ 0.15\pm0.19$ & $-0.46\pm0.19$ & B \\
  71 &  J111200.43+553908.2 &  0.4 &$ -20.0$&$     0.51  \pm   0.17$&                                       --& --& -- &  B \\
  72 &  J111202.53+554335.7 &  0.4 &$ -16.2$&$     1.56  \pm   0.36$&                                       --& --& -- &  B \\
  73 &  J111204.47+554355.7 &  1.5 &$  -8.8$&$     0.32  \pm   0.15$&                                       --& --& -- &  B \\
  74 &  J111208.84+553905.7 &  0.5 &$ -16.4$&$     0.77  \pm   0.19$&                                       --& --& -- &  B \\
  75 &  J111210.95+554136.6 &  0.6 &$ -16.7$&$     0.35  \pm   0.16$&                                       --& --& -- &  B \\
  76 &  J111220.99+554256.2 &  0.5 &$ -20.0$&$     1.91  \pm   0.33$&                            --& --& $-0.23\pm0.18$ & B \\
  77 &  J111225.58+554310.6 &  0.6 &$ -15.9$&$     1.03  \pm   0.22$&                            --& --& $-0.04\pm0.16$ & B \\
  78 &  J111226.16+553750.1 &  1.0 &$ -11.0$&$     0.39  \pm   0.17$&                                       --& --& -- &  B \\
  79 &  J111229.59+554002.4 &  0.9 &$ -16.3$&$     1.20  \pm   0.33$&                                       --& --& -- &  B \\
  80 &  J111233.80+554227.9 &  0.7 &$ -20.0$&$     2.40  \pm   0.34$&               $-0.15\pm0.18$ & --& $-0.48\pm0.14$ & B \\
  81 &  J111236.28+554329.5 &  1.3 &$ -12.5$&$     0.84  \pm   0.22$&                                       --& --& -- &  B \\
  82 &  J111240.95+553825.0 &  0.5 &$ -20.0$&$     5.69  \pm   0.62$&  $-0.47\pm0.10$ & $ 0.09\pm0.15$ & $-0.46\pm0.12$ & B \\
  83 &  J111243.11+554241.3 &  1.4 &$ -20.0$&$     1.59  \pm   0.38$&                                       --& --& -- &  B \\
\enddata
\tablecomments{ Column (1): Generic source number. (2): 
{\sl Chandra} X-ray Observatory (unregistered) source name, following the
{\sl Chandra} naming convention and the IAU Recommendation for Nomenclature
(e.g., http://cdsweb.u-strasbg.fr/iau-spec.html). (3): Position 
uncertainty (1$\sigma$) in units of arcsec. (4): The false detection probability P
that the 
detected number of counts may result from the Poisson fluctuation of the local 
background within the detection aperture [log(P) smaller than -20.0 is set 
to -20.0]. (5): On-axis (exposure-corrected) source count rate in the 
0.3-7 keV %default for aciss
%0.5-8 keV %default for acisi
%1-9 keV %default for  ACIS-I_low 
band. (6-8): The hardness ratios defined as 
${\rm HR}=({\rm H-S})/({\rm H+S})$, ${\rm HR1}=({\rm S2-S1})/{\rm S}$, 
and ${\rm HR2}=({\rm H2-H1})/{\rm H}$, 
where S1, S2, H1, and H2 are the net source count rates in the 
0.3--0.7, 0.7--1.5, 1.5--3, and 3--7~keV % for ACIS-S
%0.5--1, 1--2, 2--4, and 4--8~keV % for ACIS-I 
%1--2.5, 2.5--4, 4--6, and 6--9~keV % for ACIS-I_low 
 bands, respectively, while S and H
represent the sums, S1+S2 and  H1+H2. The hardness ratios are calculated 
only for sources with individual signal-to-noise ratios greater than 4 
in the broad band (B=S+H), and only the values with 
uncertainties less than 0.2 are included.
(9): The label ``B'', ``S'', or ``H'' mark the band in 
which a source is detected with the most accurate position, as adopted in
Column (2). 
%The label ``v'' denotes that a source is a variable.
}
  \end{deluxetable}
%====================================================
%Table 3
%====================================================
\begin{deluxetable}{lllcccc}
\tabletypesize{\footnotesize}
\tablecaption{Spectral Fitting Results of Discrete X-ray Sources \label{spectrum}}
\tablewidth{0pt}
\tablehead{
\colhead{Source} & 
\colhead{Model} &
\colhead{Spectral parameters} &
\colhead{N$_H (10^{20}{\rm~cm^{-2}})$} &
\colhead{$\chi^2$/d.o.f.} &
\colhead{$f_{\rm 0.5-10 {\rm~keV}}$} &
\colhead{$f_{\rm 0.01-100 {\rm~keV}}$}}
\startdata
\noalign{\smallskip}
25\dotfill&Power law &$\Gamma =1.74(1.53-1.99)$& $\lesssim 5.7$ &13.5/19&4.7&14\\
26\dotfill&Diskbb    &$kT_{in}=1.32(1.21-1.46)$&14(11-18)&59.6/52&19&21\\
35\dotfill&Power law &$\Gamma =1.81(1.56-2.05)$&30(21-47)&38.5/31&8.2&24\\
\noalign{\medskip}
\enddata
\tablecomments{$\Gamma$ is the photon index, whereas $kT_{in}$ (in units
of keV) is the
effective inner temperature of the accretion disk. 
The uncertainty ranges of the parameters are included in
the parentheses and are all at the 90\% confidence. The unabsorbed 
flux $f$ is in units of $10^{-14}{\rm~erg~s^{-1}~cm^{-2}}$}.
\end{deluxetable}

%====================================================
%Table 4
%====================================================
\begin{deluxetable}{lr}
\tabletypesize{\footnotesize}
\tablecaption{Diffuse X-ray Spectral Fit}
\tablewidth{0pt}
\tablehead{
\colhead{Parameter} & 
\colhead{Value}}
\startdata
\noalign{\smallskip}
Best-fit $\chi^2$/d.o.f.\dotfill  &                     43.8/35 \\
\noalign{\medskip}
Low temperature (keV)\dotfill                    &      0.166(0.158-0.175)\\
Integrated EM   ($10^{10}~{\rm cm}^{-5}$) \dotfill &    1.2(0.5-2.5) \\ 
High temperature (keV)\dotfill                   &      0.62(0.58-0.66)\\
Integrated EM   ($10^{10}~{\rm cm}^{-5}$) \dotfill &    0.6(0.3-1.2) \\ 
Abundances (solar)\dotfill  & 0.33 (0.13-1.6)\\
Column Density  ($10^{20}~{\rm cm}^{-2}$)\dotfill  &    2 (fixed) \\
\enddata
\tablecomments{The parameters are from the two-temperature thermal plasma model
fit to the diffuse X-ray spectrum of NGC 3556 (see also Fig. 9).
The uncertainty ranges of the parameters are included in
the parentheses and are all at the 90\% confidence.
The integrated emission measure is defined as $1/4\pi \int_{\Omega} {\rm EM} d\Omega$,
where the integration is over a solid angle over which the data were collected and
${\rm EM}=\int n_e^2 dr$ with $n_e$ being the electron density (all units in cgs).}
\end{deluxetable}

%====================================================
%Table 5
%====================================================
\begin{deluxetable}{l c c c c}
  \tabletypesize{\footnotesize}
  \tablecaption{X-Ray and Radio Source Correlations}
  \tablewidth{0pt}
  \tablehead{
  \colhead{Source} &
  \colhead{Radio Flux (mJy)} &
  \colhead{Radio FWHM($^{\prime\prime}$)} &
  \colhead{Radio S/N} &
  \colhead{Offset ($^{\prime\prime}$)}
  }
 \startdata
  18 & 0.72 $\pm$ 0.06 & 2.4 $\pm$ 0.6& 34.5 &  0.6 $\pm$ 0.4 \\
  24 & 0.30 $\pm$ 0.05 & 2.0 $\pm$ 0.3& 14.4 &  1.1 $\pm$ 0.7 \\
  36 & 2.13 $\pm$ 0.07 & 2.8 $\pm$ 0.6& 101.5&  0.6 $\pm$ 0.5 \\
  53 & 0.42 $\pm$ 0.05 & 2.1 $\pm$ 0.4& 20.0 &  0.7 $\pm$ 0.3 \\
 \enddata
 \tablecomments{Both the flux and the full width at half maximum (FWHM) 
of radio emission are calculated from a Gaussian fit to the 20 cm continuum 
data (Irwin et al. 2000, beam size =
1\farcs7 $\times$ 1\farcs4)
de-convolved from the beam. 
The signal-to-noise ratios (S/N) are 
computed by taking the ratio of the flux to the rms noise 
(0.021 mJy/beam) of the radio data. The offset is between the X-ray 
source position (Column 2 of Table 2) and the centroid position of 
the Gaussian fit. }
\end{deluxetable}

%========================================================================
%Figures
%========================================================================
\begin{figure} 
\centerline{
\psfig{figure=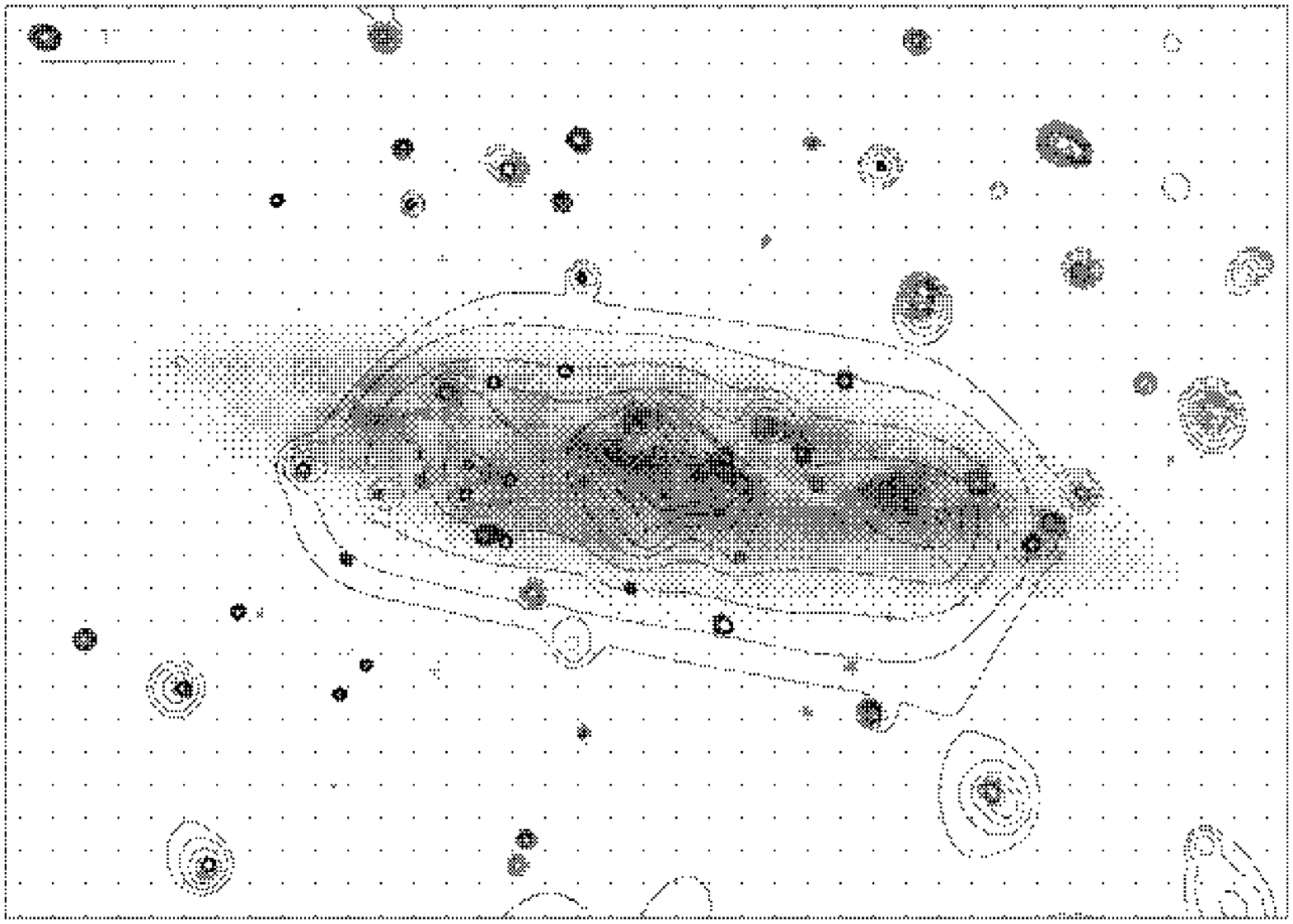,height=2.7in,angle=90, clip=}
\psfig{figure=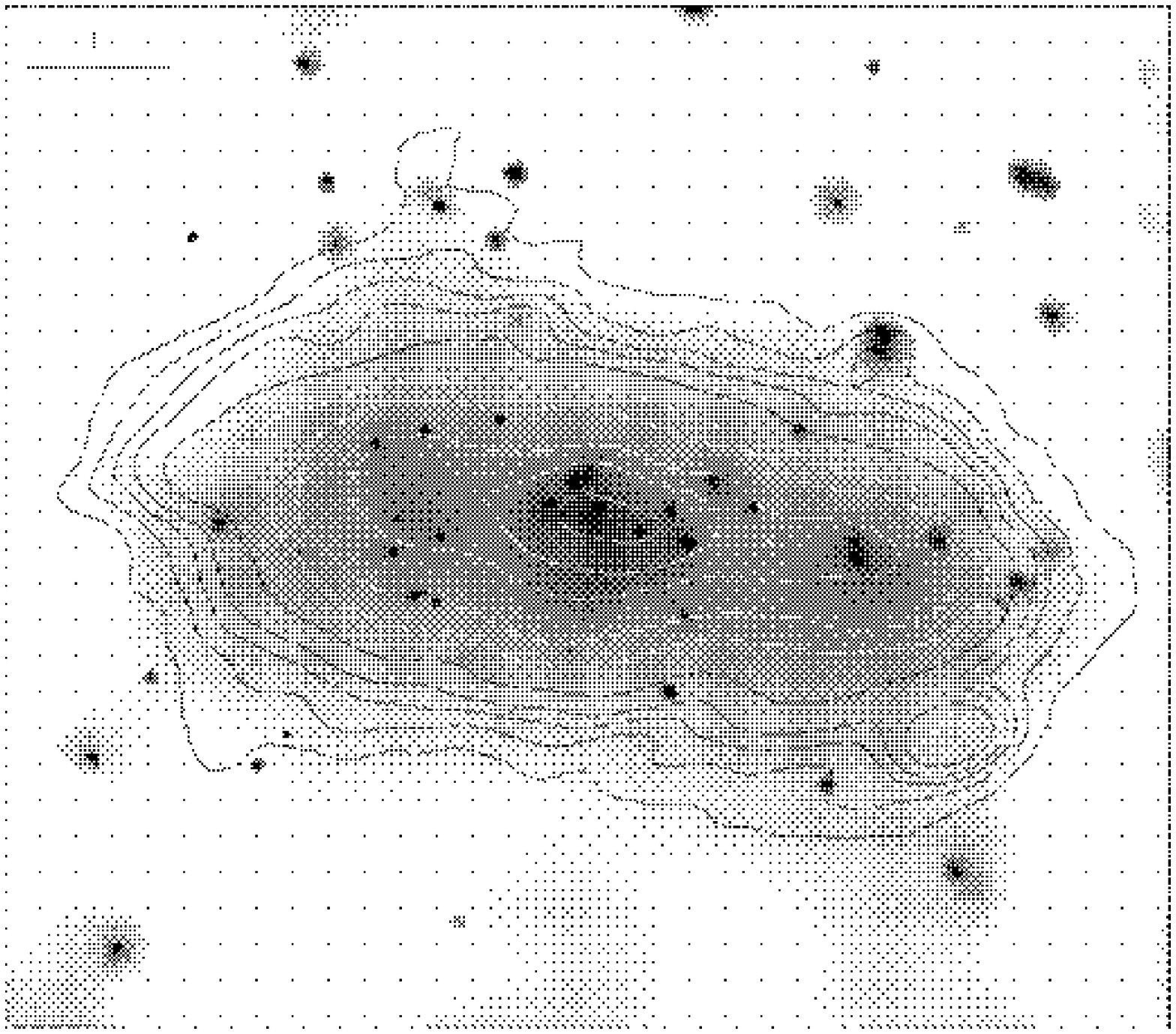,height=2.7in,angle=90, clip=}
}
\caption{{\sl Left-hand panel:} \xs~ in optical blue (gray-scale; DSS)
and in the ACIS-S 0.3-7 keV band. The X-ray intensity image is 
adaptively smoothed 
to achieve a S/N ratio of $\sim 3$ (as defined in the CIAO program CSMOOTH).
The contours are at 6.4, 6.8, 7.5, 8.6, 10, 12, 14, 16,
23, 41, and 77 $\times 10^{-3} {\rm~ct~s^{-1}~arcmin^{-2}}$. 
{\sl Right panel:}  \xs~ in the 0.3-1.5 keV band (gray-scale) and in 20 cm
continuum (contours; Irwin et al. 1999). The X-ray image is smoothed with
the exactly same kernel as that in the left-hand panel. The radio contours 
are at 2, 4, 6, 8, 12
20, 30, 60, 100, and 200 $\times 10^{-4} {\rm~Jy~beam^{-1}}$. }
\end{figure}

\begin{figure} 
\centerline{
\psfig{figure=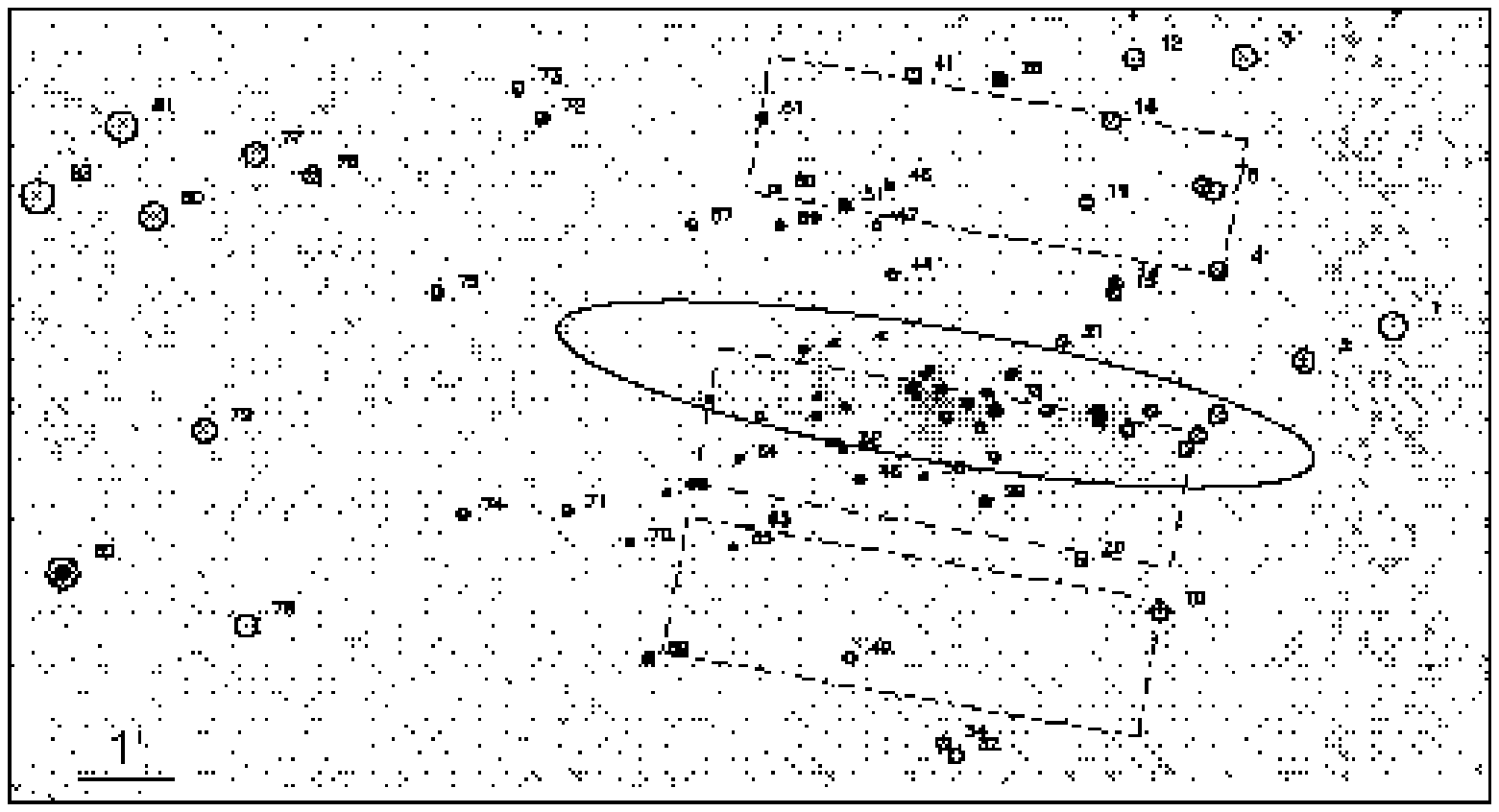,height=3.6in,angle=90, clip=}
}
\centerline{
\psfig{figure=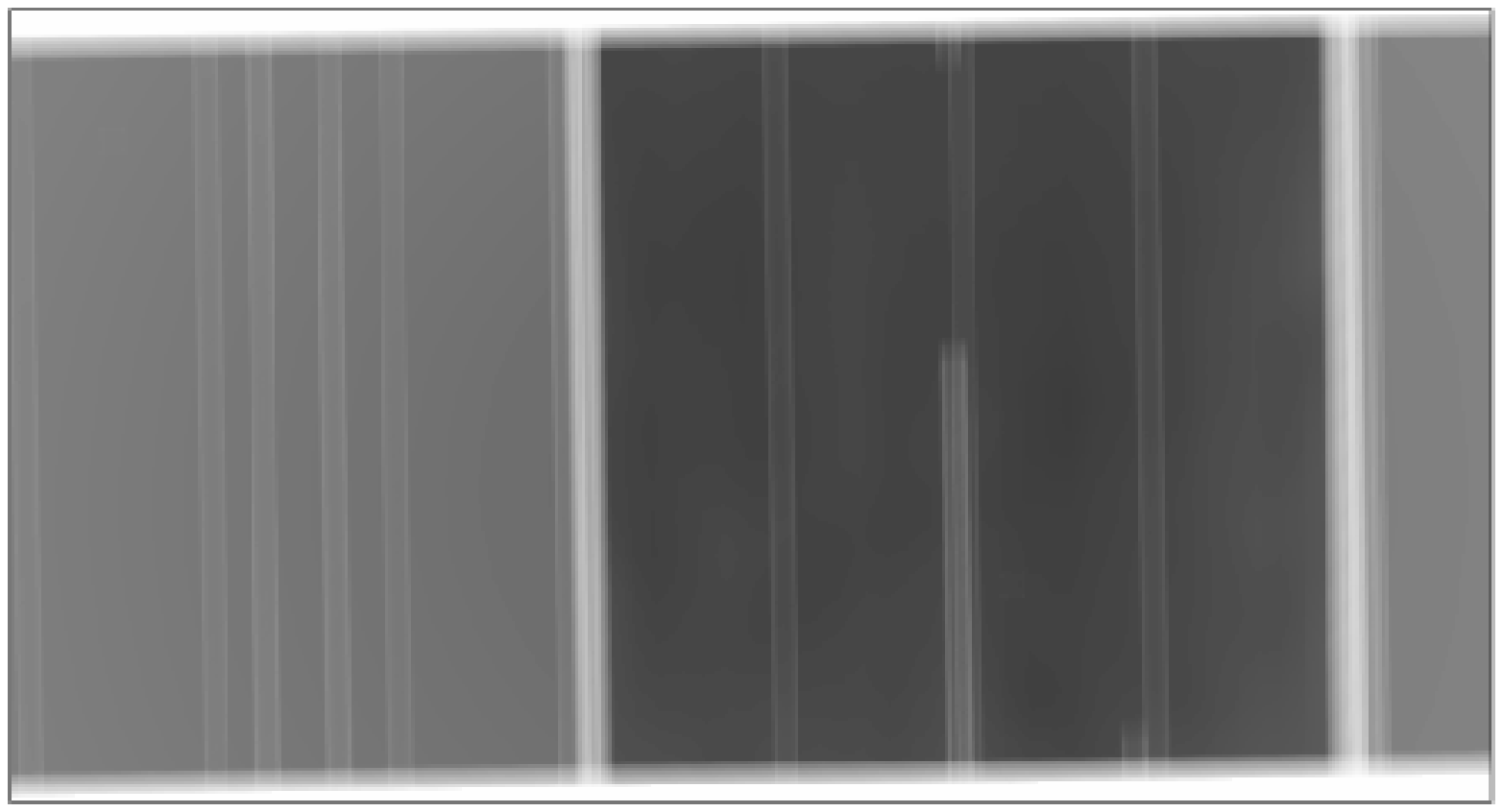,height=3.6in,angle=90, clip=}
}
\caption{ACIS-S data of NGC 3556. The upper panel shows 
a 0.3-7 keV band intensity image, which is
smoothed with a Gaussian of FWHM equal to 3\farcs9. The circles enclose
the source regions which are defined in the text. The radius of
a source depends on its off-axis location and is smaller than 
its {\sl apparent} size near the axis in this smoothed image.
Generic source numbers (Table 2) are marked for sources located outside the 
$R_{25}$ ($I_B = 25$ mag arcsec$^{-2}$ isophote) 
ellipse of the galaxy. The rectangular boxes
illustrate the regions from which the on-galaxy diffuse X-ray spectrum 
(the dashed box) and the off-galaxy background spectrum (the two dash-dotted 
boxes; Fig. 9).
As an example, the lower panel shows an exposure map in the 0.7-1.5 keV band
in the same field.
}
\end{figure}
\begin{figure} 
\centerline{ 
\psfig{figure=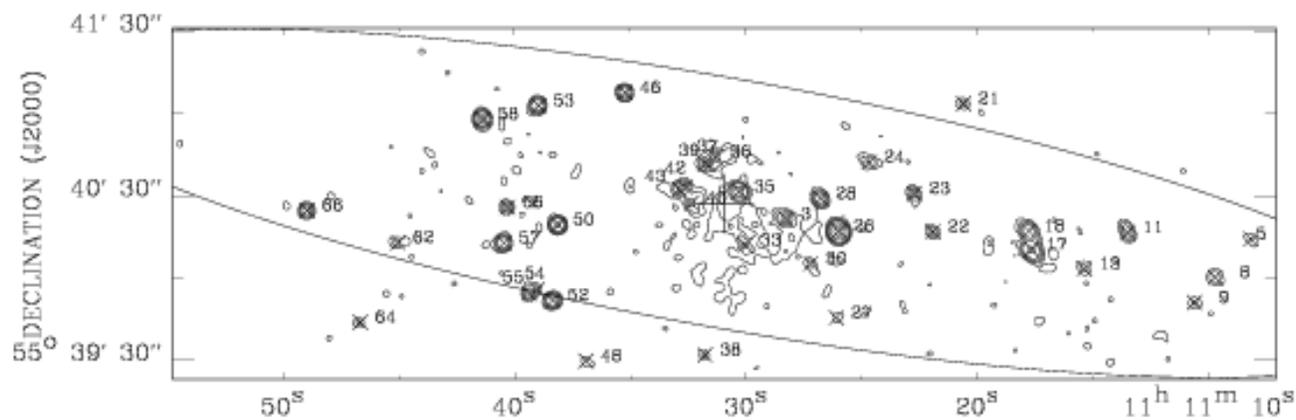,height=2.2in,angle=90, clip=}
}
\caption{A close-up of the X-ray intensity image of NGC 3556 (Fig. 2
upper panel).
The image is smoothed with a Gaussian of FWHM equal to 3$^{\prime\prime}$.  
The contours are at  3,  6, 12, 25, 100, 400, and 1000
 $\times 10^{-2} {\rm~count~s^{-1}~arcmin^{-2}}$. 
Source positions and numbers are marked by crosses (see also Table 1).
The optical center of the galaxy is marked by the
large plus sign.}
\end{figure}

\begin{figure} 
\centerline{ 
\psfig{figure=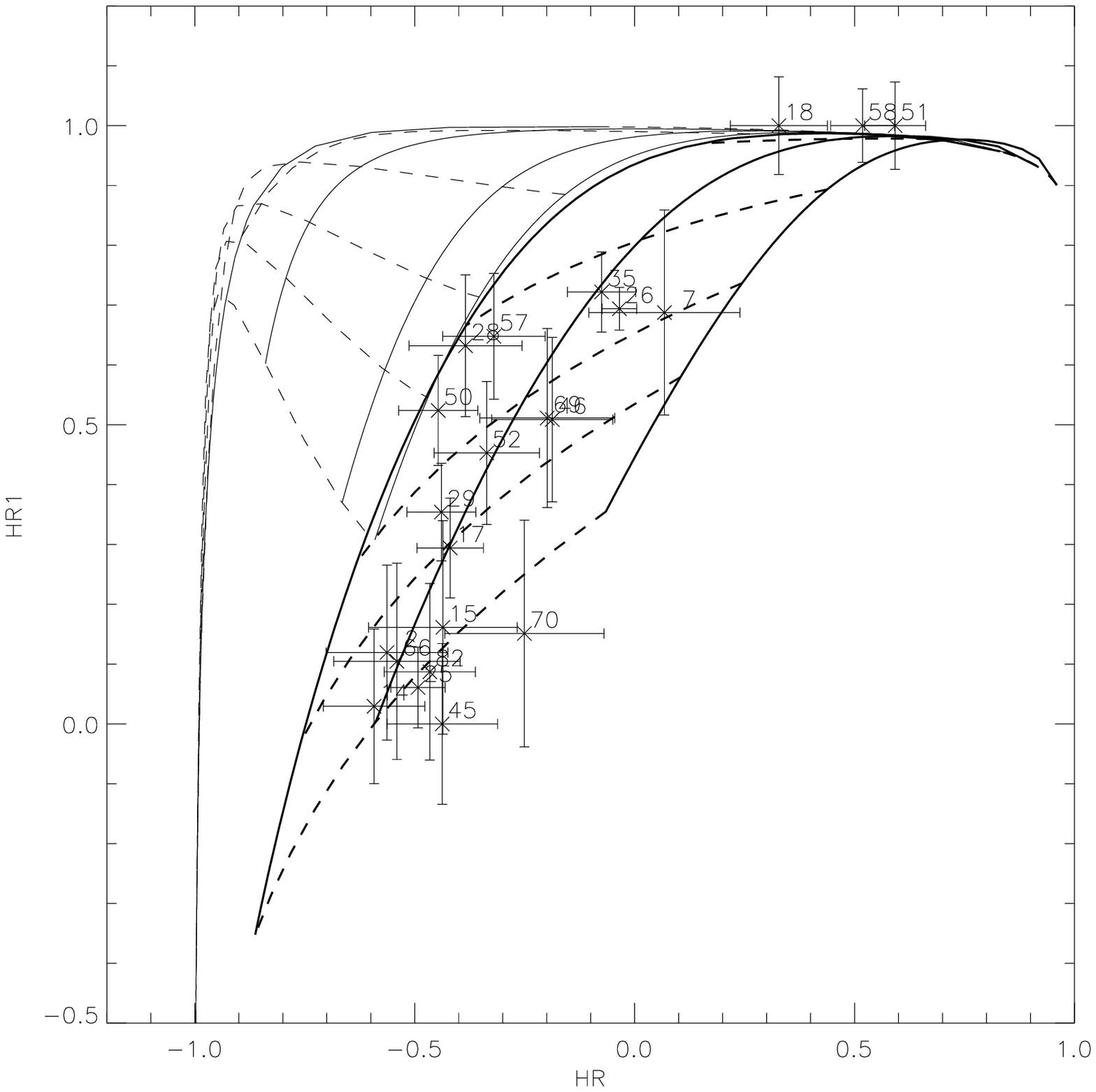,height=3.3in,angle=90, clip=}
\psfig{figure=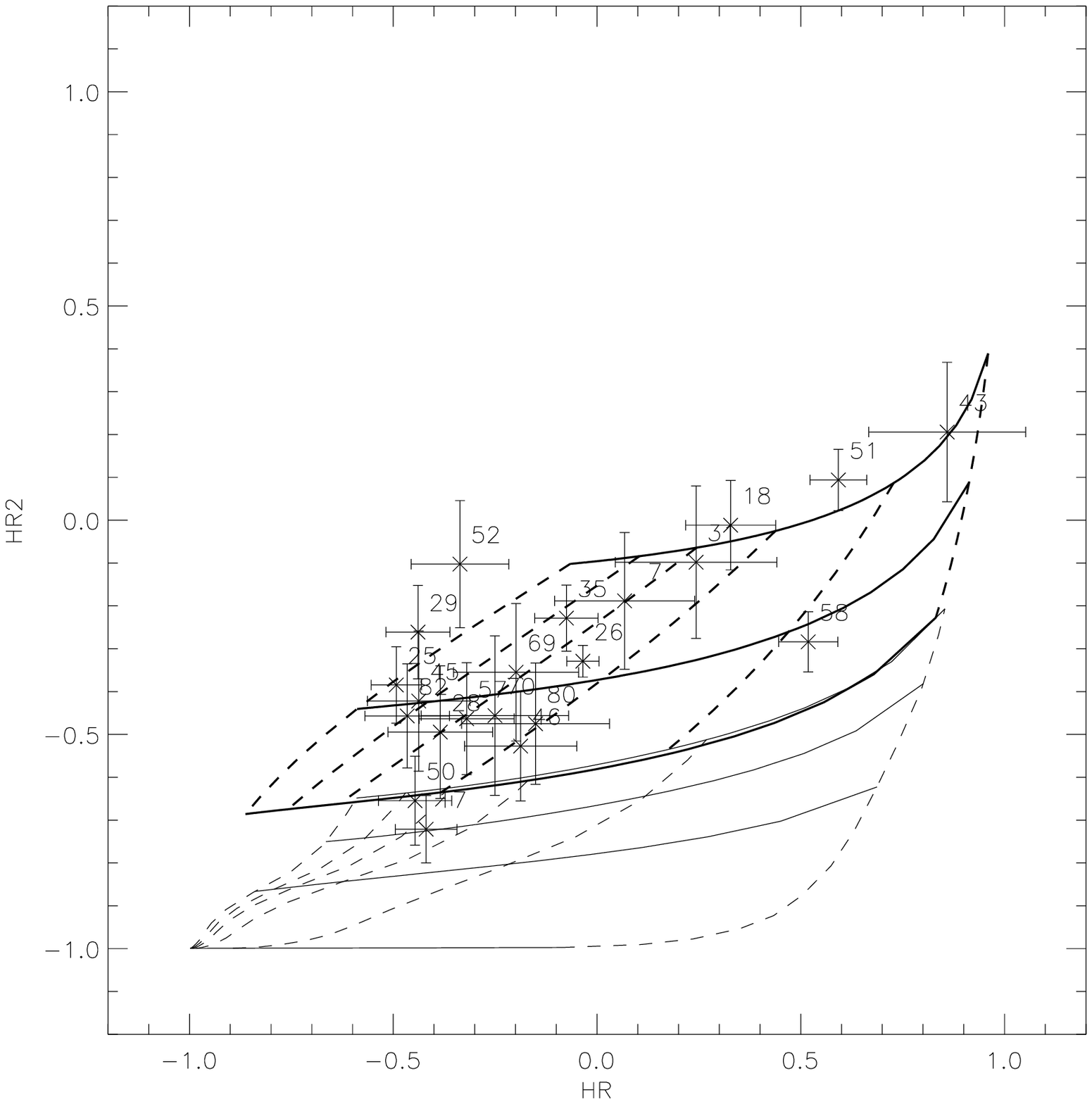,height=3.3in,angle=90, clip=}
}
\caption{Color-color diagrams of X-ray sources. The hardness ratios (HR1,
HR2, and HR) and their 1$\sigma$  error bars are defined in the notes 
to Table 2.
The generic source numbers (Table 2) are marked. Also included in the plot are
hardness-ratio models: the solid thick curves are for the power-law model
with a photon index equal to 1, 2, and 3, whereas the 
solid thin curves are for the thermal plasma with a temperature
equal to 0.2, 1, 1.5, and 2 keV, from right to left in the left 
panel and from top to bottom in the right panel, respectively. The absorbing
gas column densities are 1, 10, 20, 40, 100, and 300
$\times 10^{20} {\rm~cm^{-2}}$ (dashed curves from bottom to top in the left 
panel and from left to right in the right panel). 
}
\end{figure}

\begin{figure}
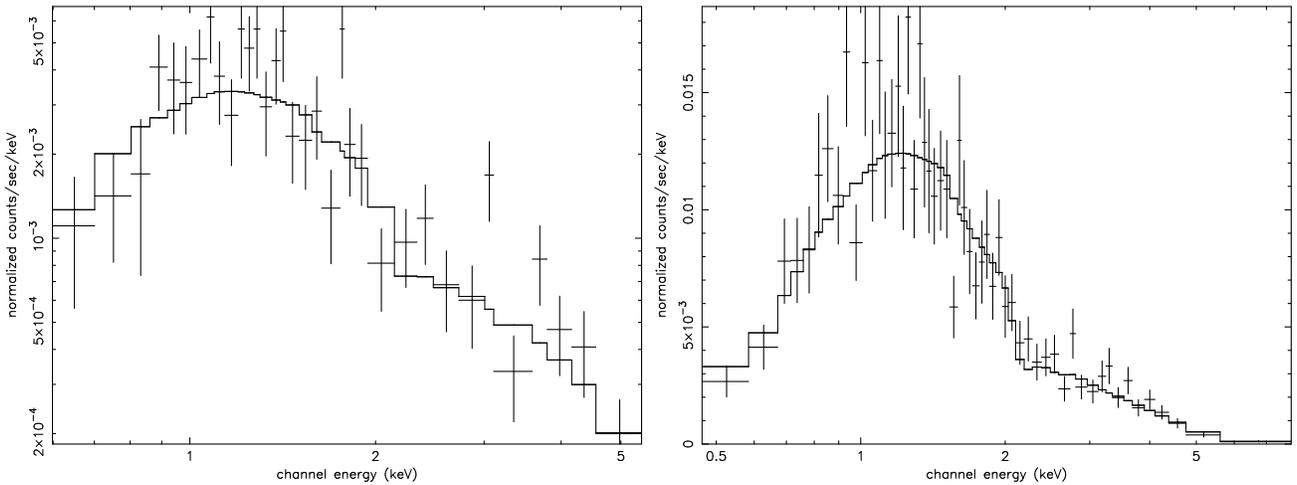
 
\centerline{ 
\psfig{figure=f5a.ps,height=2.5in,angle=270, clip=}
\psfig{figure=f5b.ps,height=2.5in,angle=270, clip=}
}
\caption{ACIS-S spectra of Sources 35 (left-hand panel) and 25 
(right-hand panel). The histogram in each plot represents the best 
fit model spectrum as listed in Table 3.
}
\end{figure}

\begin{figure} 
\centerline{ 
\psfig{figure=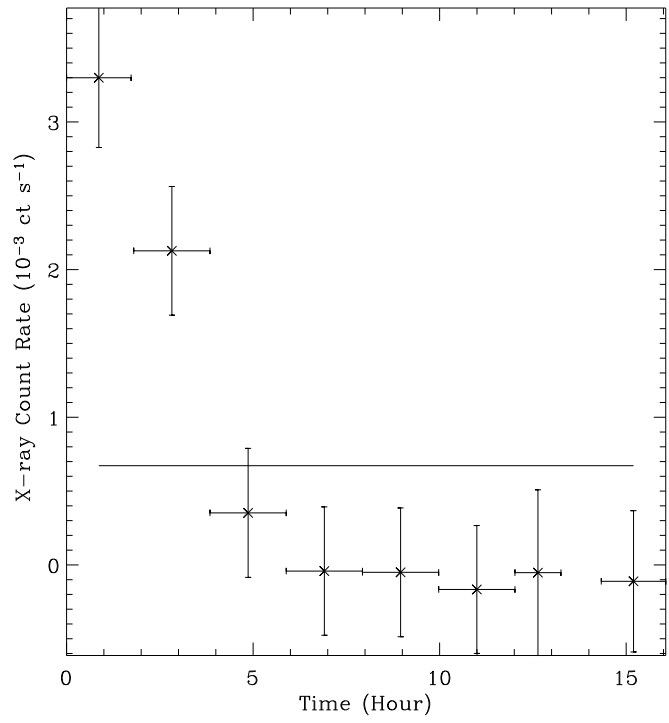,height=4in,angle=90, clip=}
}
\caption{Light-curve of Source 31. The horizontal line represents the
mean count rate of the source.
}
\end{figure}

\begin{figure} 
\centerline{ 
\psfig{figure=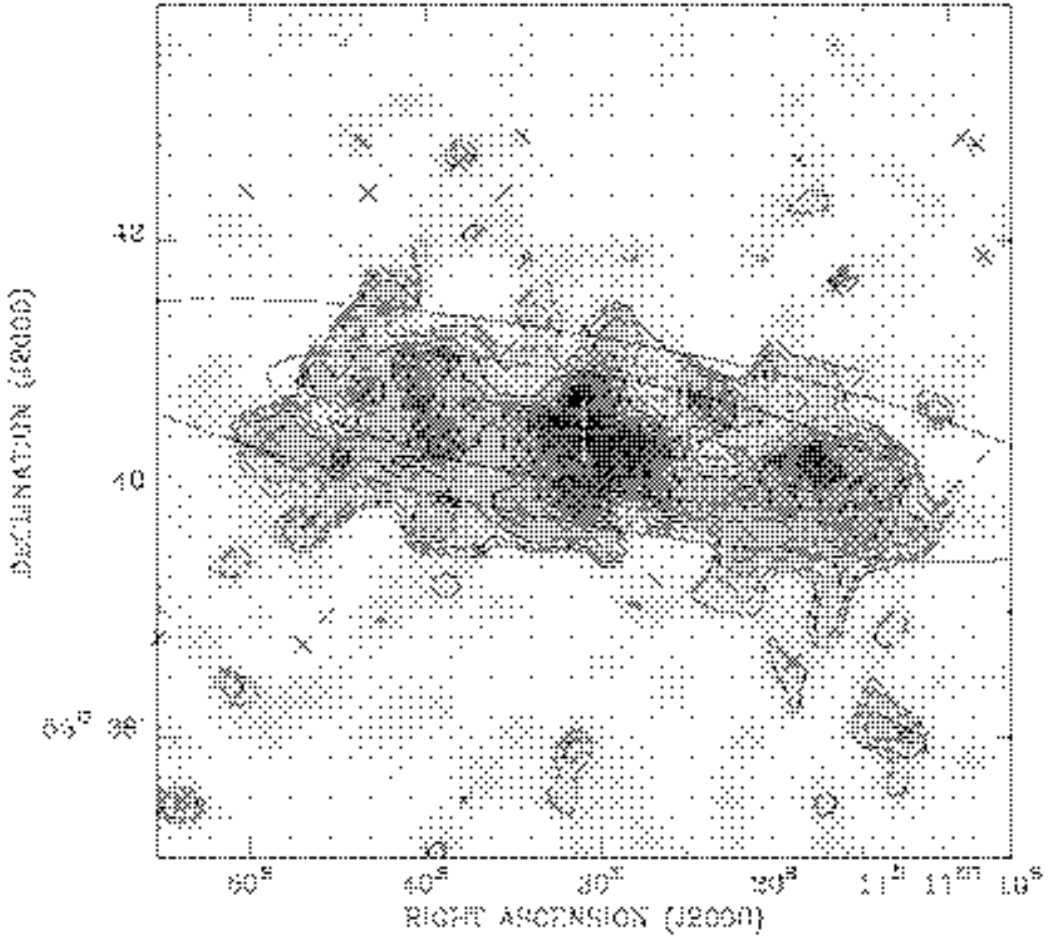,height=6in,angle=90, clip=}
}
\caption{Diffuse emission from NGC 3556 in the 0.3-1.5 keV
band. The image is smoothed adaptively with a
Gaussian, the size of which is adjusted to achieve a count-to-noise 
ratio of 6. The contours are at 3, 4, 6, 9, 13, 18, 24, 31, 50, and 100 
$\sigma$ above the background ($3.5 \times 10^{-3} 
{\rm~ct~s^{-1}~arcmin^{-2}}$;
1$\sigma = 0.59 \times 10^{-3} {\rm~ct~s^{-1}~arcmin^{-2}}$).
The location of the sources, removed from the data before
the smoothing, are marked by crosses. Ellipses of  $R_{25}$ and 
3/5  $R_{25}$ are included for comparison with the extent of the diffuse
X-ray emission. The optical center of the galaxy is also marked by the
large plus sign.
}
\end{figure}

\begin{figure} 
\centerline{ 
\psfig{figure=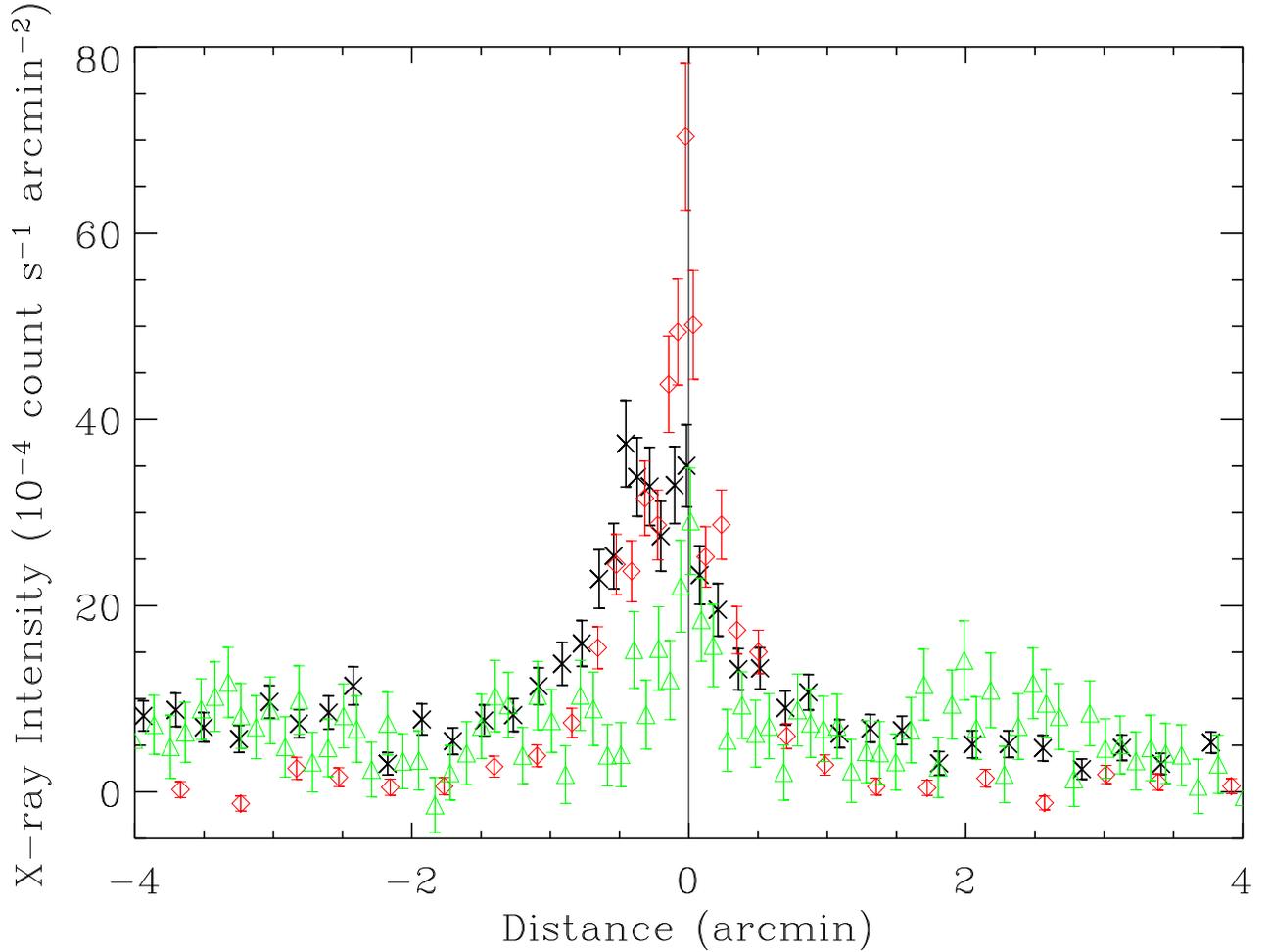,height=5.4in,angle=0, clip=}
}
\caption{ACIS-S intensity distribution across the galactic disk in 
0.3-0.7 keV (crosses), 0.7-1.5 keV (diamonds), 
and 1.5-7 keV (triangles) bands. The width
used for the intensity average is fixed to be 6/5 $R_{25}$ (Fig. 7),
symmetric relative to the minor axis.
 The step of the average in the direction of
the minor axis is adaptively adjusted to achieve
a count-to-noise ratio of 10. The blank-field background, calculated using
the same step, has been subtracted from the data points. 
The vertical straight line in the middle represents the position of the major 
axis of the galaxy, whereas the horizontal axis 
marks the distance from the major axis of the galaxy
(southeast on the left).
}
\end{figure}

\begin{figure} 
\centerline{ 
\psfig{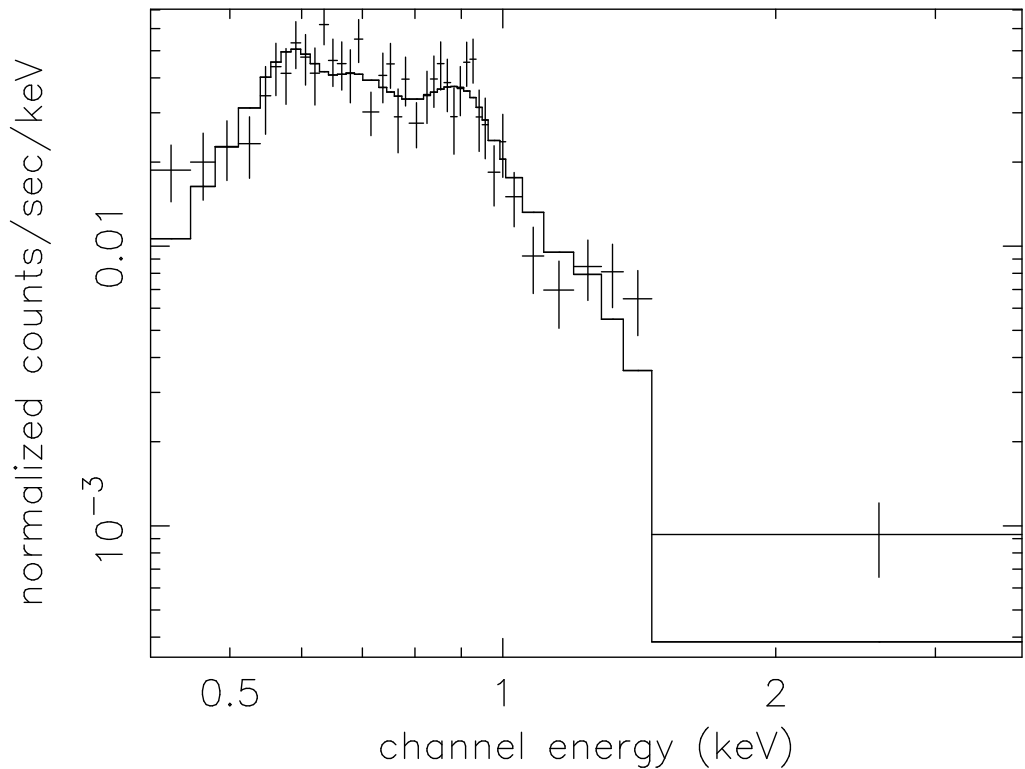}
}
\caption{ACIS-S spectrum of diffuse X-ray emission from NGC~3556. 
The on-galaxy and off-galaxy (subtracted) background spectral 
data were extracted from rectangular boxes illustrated in
Fig. 2 (upper panel). The histogram represents the best fit
with a two-temperature thermal plasma (MEKAL) model. The spectral parameters
are summarized in Table 4.
}
\end{figure}

\begin{figure} 
\centerline{ 
\psfig{figure=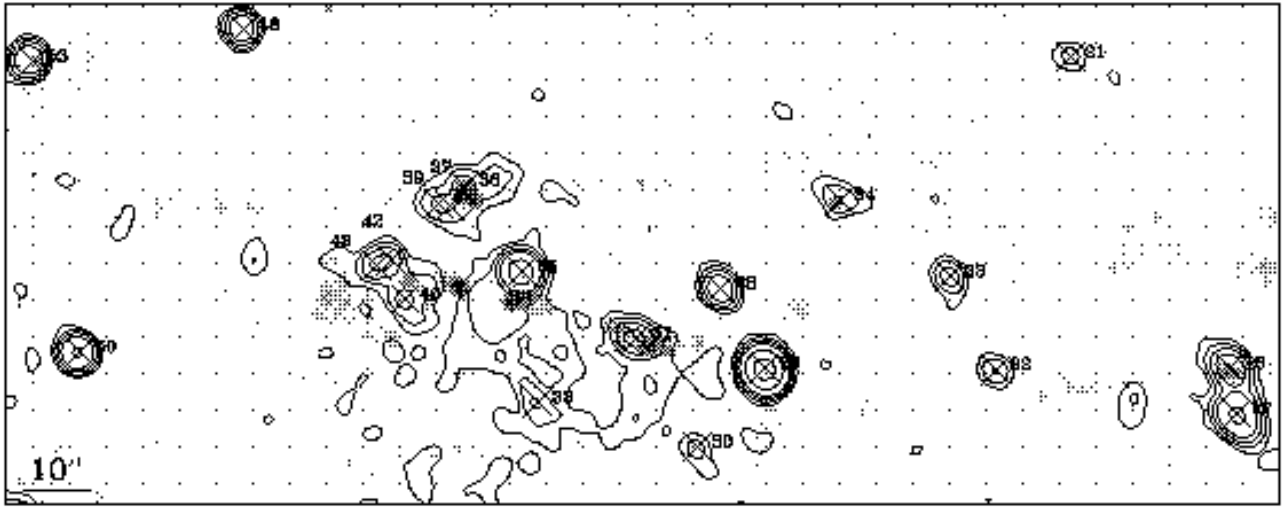,height=2.7in,angle=90, clip=}
}
\caption{Comparison of ACIS-S 0.3-7 keV intensity contours  (Fig. 3)
overlaid on 
radio continuum image (20~cm; gray-scale). The X-ray sources
are numbered as in Column~1 of Table 2.  
The resolution of the radio data is 1\farcs8 and the rms noise is 
$0.021 {\rm~mJy~beam^{-1}}$ (Irwin et al, 2000).
}
\end{figure}

\begin{figure} 
\centerline{ 
\psfig{figure=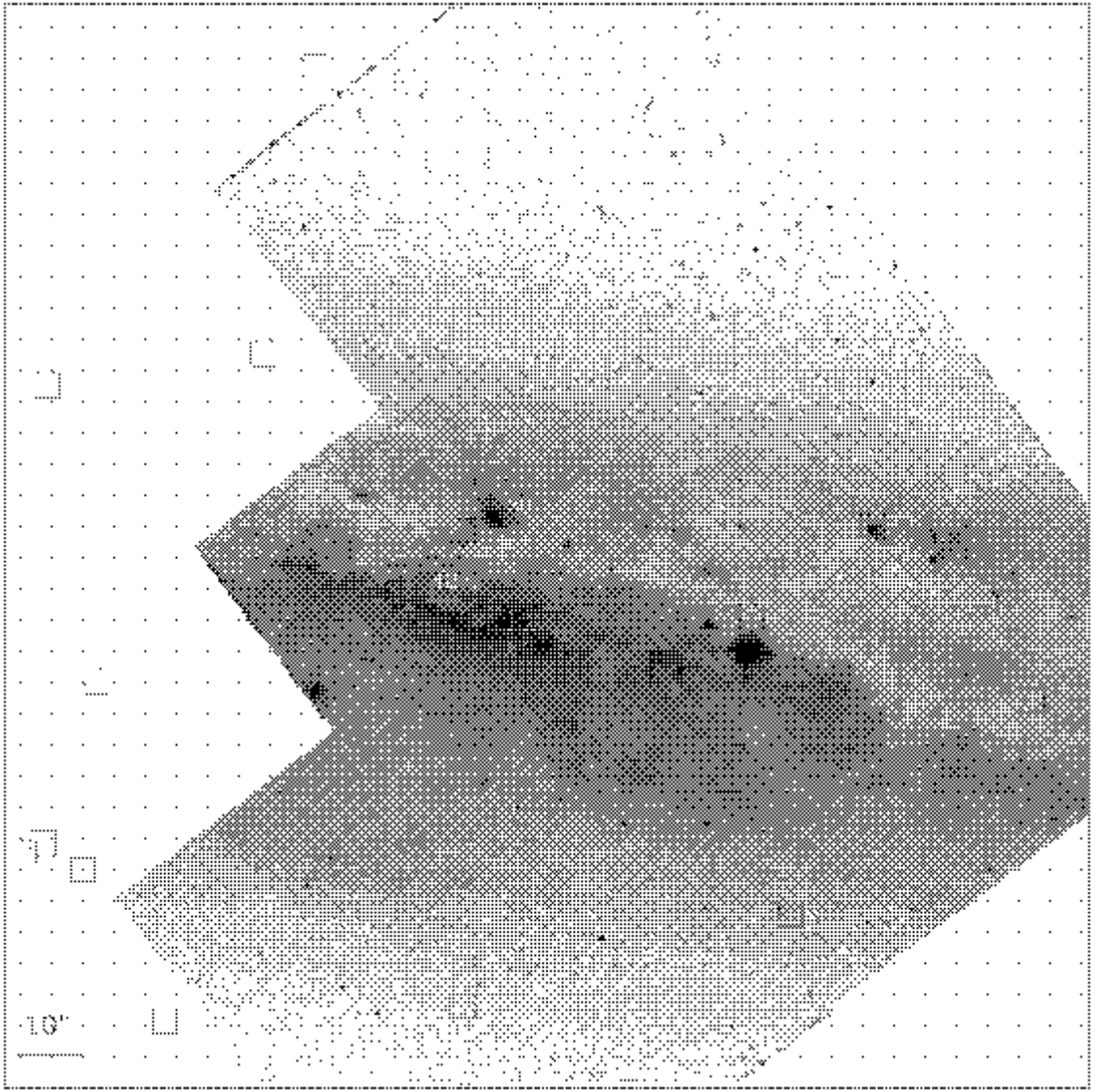,height=6in,angle=90, clip=}
}
\caption{{\sl HST} WFPC2 image, compared with the 
diffuse X-ray emission contours covering the central region of NGC 3556. 
The contour levels are the same as in Fig. 7. X-ray sources (Table 2;
Fig. 3) are marked as squares. In particular, the three sources with possible
optical counterparts are labeled. The center of the galaxy is marked
as a plus sign.
}
\label{fig:HST}
\end{figure}

\begin{figure} 
\centerline{ 
\psfig{figure=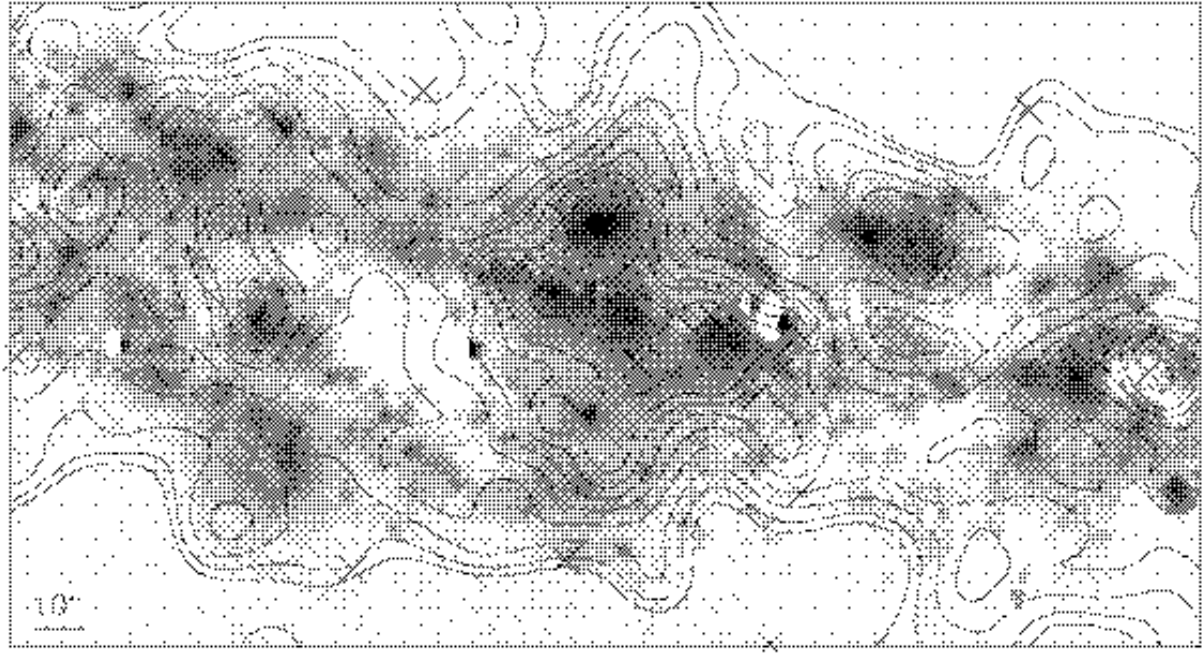,height=3.5in,angle=90, clip=}
}
\caption{Comparison of the diffuse X-ray emission contours (Fig. 7) and 
the H$\alpha$ image (Collins et al. 2000; Artifacts from imperfect continuum
subtraction are apparent near bright objects).  
X-ray point source positions are marked with crosses.
}
\end{figure}

\end{document}